\documentclass[sn-basic]{sn-jnl}

\jyear{2021}%

\theoremstyle{thmstyleone}%
\theoremstyle{thmstyletwo}%

\usepackage{bm}
\usepackage{cleveref}

\theoremstyle{thmstylethree}%

\begin{document}

\title[Modelling installation noise control]{Modeling closed-loop control of installation noise using Ginzburg-Landau equation}


\author*[1]{\fnm{Ugur} \sur{Karban}}\email{ukarban@metu.edu.tr}

\author[2]{\fnm{Eduardo} \sur{Martini}}\email{eduardo.martini@ensma.fr}

\author[2]{\fnm{Peter} \sur{Jordan}}\email{peter.jordan@univ-poitiers.fr}

\affil*[1]{\orgdiv{Aerospace Engineering Department}, \orgname{Middle East Technical University}, \orgaddress{ \city{Ankara}, \postcode{06800}, \country{Turkey}}}

\affil[2]{\orgdiv{D\'{e}partement Fluides, Thermique, Combustion}, \orgname{Institut PPrime, CNRS–Universit\'e de Poitiers–ENSMA}, \orgaddress{\city{Poitiers}, \postcode{86036}, \country{France}}}



\abstract{Installation noise is a dominant source associated with aircraft jet engines. Recent studies show that linear wavepacket models can be employed for prediction of installation noise, which suggests that linear control strategies can also be adopted for mitigation of it. We present here a simple model to test different control approaches and highlight the potential restrictions on a successful noise control in an actual jet. The model contains all the essential elements for a realistic representation of the actual control problem: a stochastic wavepacket is obtained via a linear Ginzburg-Landau model; the effect of the wing trailing edge is accounted for by introducing a semi-infinite half plane near the wavepacket; and the actuation is achieved by placing a dipolar point source at the trailing edge, which models a piezoelectric actuator. An optimal causal resolvent-based control method is compared against the classical wave-cancellation method. The effect of the causality constraint on the control performance is tested by placing the sensor at different positions. We demonstrate that when the sensor is not positioned sufficiently upstream of the trailing edge, which can be the case for the actual control problem due to geometric restrictions, causality reduces the control performance. We also show that this limitation can be moderated using the optimal causal control together with modelling of the forcing. }




\maketitle

\section{Introduction}\label{sec:intro}

Jet noise is becoming an increasingly important issue for the aviation industry. The reduced exit velocity in modern ultra-high-by-pass-ratio (UHBR) turbofan engines helped significantly reduce the direct noise emitted by jet engines. However, the increased nozzle diameter in these engines resulted in a smaller distance between the jet lip line and the wing structure, which caused an increase in the noise due to the interaction between the jet and the flaps on the wing, called the \emph{installation noise}. In this study, we focus on modelling and control of the installation noise problem.

The noise generation mechanisms associated with installed jets, i.e., jets in the vicinity of a wing, are known to be different from that in free jets. In both cases, however, wavepackets constitute a key element. For free jets at subsonic speed, the jitter of wavepackets are shown to be an important noise source, as it causes a coherence decay in wavepackets, increasing their acoustic efficiency \citep{cavalieri_jsv_2011,cavalieri_jfm_2014,cavalieri_amr_2019}. For a wavepacket model to accurately predict jet noise, it should exhibit similar coherence decay as in a real jet. This implies that a rank-1 wavepacket model, which has unit coherence across the domain by construction, cannot yield an accurate noise prediction even if it decently captures the turbulent fluctuations in the shear layer. This is the case for installed jets as well if the distance between the nozzle exit and the trailing edge (TE) of the flap is large. It was shown in \citet{nogueira_aiaa_2016} that a model for coherence decay was necessary for accurate noise prediction with four diameters separation between the nozzle and the TE. Contrary to their result, \citet{faranosov2019modeling} showed that using a wavepacket model obtained by solving linear parabolised stability equations, which is rank-1 by construction, it was possible to predict the installation noise with good accuracy when the TE is located about three diameters downstream the nozzle. This suggests that, as the separation between the TE and the nozzle exit becomes shorter, a rank-1 wavepacket model may be sufficient for noise prediction. A distance shorter than three diameters constitutes a more relevant case as in many published scaled-model tests with UHBR turbofan engines \citep{davy_aiaa_2019}, the distance between the nozzle exit and flap TE is about two-to-three diameters. 

Once an appropriate source model is available, noise generation in installed jets can be modelled using aeroacoustic analogies. The direct noise emitted by the source, which is called incident noise, can be predicted using Lighthill's acoustic analogy \citep{lighthill1952sound}. \citet{fwilliams_jfm_1970} introduced a tailored Green's function (TGF), which can account for the scattered acoustic field due to a source near a semi-infinite plate using Curle's analogy \citep{curle_rspa_1955}. There exist many studies \citep{cavalieri2014scattering,nogueira_aiaa_2016,piantanida2016scattering,dasilva_jasa_2019} showing that the TGF for semi-infinite plates can be used for accurate prediction of the acoustic field in an installed jet configuration. 

All these developments reported in the literature point to the hypothesis that installation noise can be modelled as a linear mechanism involving a linear source model and a TGF for noise generation and propagation, respectively. This opens the possibility of applying a linear control for installation noise via real-time measurement of turbulent fluctuations at a single station around the nozzle exit, to predict the wavepacket which is to be formed downstream by convection and amplification of these measured fluctuations. Once an estimate of the wavepacket is available, the noise emitted by this wavepacket can be predicted using a TGF. One, of course, needs a proper actuator capable of producing an acoustic field similar to that of the installed jet configuration, to be able to control the installation noise. The acoustic directivity of installed jets is known to be of the dipolar type, due to the strong amplification of sound by the TE, acting as a point dipole. 

Installation noise control is a growing research topic. There exist a number of studies which investigate passive control strategies for installation noise. \citet{rego_jsv_2021} and later \citet{jente_aiaa_2022} investigated the potential of using permeable surfaces to reduce installation noise. In a similar approach, \citet{jawahar_jsv_2023} studied porous plates to achieve installation noise reduction. \citet{mancinelli_aiaa_2022} investigated using flexible surfaces at the TE of a solid plate, rendering the TE permeable to the pressure fluctuations induced by the wavepacket. 

Studies on active control of installation noise, on the other hand, are more scarce. To the best of the authors' knowledge, the literature is limited to two studies \citep{kopiev2019plasma,kopiev2020active}, where they investigate the use of plasma actuators for controlling instabilities in jets, leading to a reduction in the installation noise. In this study, we explore the potential of an alternative active control strategy based on actuation at the TE of a solid surface. We adopt a closed-loop control strategy similar to the ones used in \citet{maia2020closed} and \citet{maia2021real} for control of free jets. We use an ideal model problem, which contains a linear wavepacket near a semi-infinite plate and a point dipole attached to the TE of the plate. The linear wavepacket is obtained by solving the Ginzburg-Landau equation with stochastic external forcing. The forcing, the parameters of the G-L problem and the position of the semi-infinite plate are configured such that the resulting acoustic field is reminiscent of that in an actual low subsonic jet. The stochastic fluctuations in the wavepacket are measured using a point sensor to determine the control input. We investigate the effect of causality constraint on the control performance in a realistic configuration where the distance between the sensor and the TE is similar to that between the nozzle exit and the TE of the flap in an actual aircraft. We aim at finding the optimal sensor position for control and investigate the feasibility of such a control action given the geometric restrictions in an actual set-up on aircraft even if one has a perfectly linear noise generation mechanism and an ideal actuator for control. 

We implement two different control strategies: the wave-cancellation method and the optimal causal resolvent-based control. Both are frequency-domain approaches. The former is a classical linear control method that involves computing linear transfer functions between the three elements of linear control: sensors, observers and actuators, and finding a control law to cancel the fluctuations at the observer. It has been used for flow control in many studies \citep{thomas_jfm_1983,laurien_jfm_1989,li_jfm_2006,sasaki2018closed,sasaki2018wave,
maia2021real}. As the control law is, in general, non-causal, it is often necessary to truncate it to its causal part. The latter approach has been proposed by \citet{martini_jfm_2022}, where causality is imposed as a constraint while calculating the optimal control kernel. This approach was seen to provide significantly better performance than the wave-cancellation approach when the distance between sensors and actuators is small. Enforcing causality in the frequency domain is achieved via Wiener-Hopf decomposition \citep{youla_ieee_1976}. The method was tested in \citet{martini_jfm_2022} to control turbulent fluctuations in an amplifier flow, similar to the case of a jet. We compare the two methods at varying sensor positions to verify the above-mentioned hypothesis.

The outline of the paper is as follows: we revisit the two control methods in section \ref{sec2}. We give the details of the Ginzburg-Landau model and the TGF function for semi-infinite plate in section \ref{sec:linmod}. We present the results of implementing the control methods on the model problem in section \ref{sec:imp}. And, we conclude the paper in section \ref{sec:conc} with some final remarks.

\section{Control methods}\label{sec2}
The model used in this study contains two parts: (\emph{i}) the wavepacket model, which can be represented as a 1D linear time-domain system as
\begin{align} \label{eq:linsys}
\partial_t q(x,t) -\mathcal{A}q(x,t) = f(x,t),
\end{align}
where $q$ and $f$ denote the state and the external forcing and $\mathcal{A}$ denotes the linear time-invariant operator; and (\emph{ii}) the acoustic propagation problem solved by the tailored Green's function (TGF) used for calculating the sound generated by the wavepacket $q(x,t)$. 

Optimal causal control for spatially distributed systems can be achieved by a linear quadratic regulator (LQR) \citep{motee_ieee_2008}. Such a controller requires a state-space representation of the system, including the acoustic field. Contrary to the wavepacket model, the TGF, which takes into account the acoustic scattering due to the semi-infinite plane, is only available in the frequency domain. Obtaining a dynamic time-domain model for the acoustic propagation problem then becomes a non-trivial task. Besides, formulating a state-space representation including the acoustic field renders the cost of the LQR problem dependent on the observer position. It is customary to place the observers far from the source domain, which can make the problem too large for LQR-based methods. Given these limitations, we focus on control methods which are based on transfer functions between sensors, actuators and targets. 
\subsection{Wave-cancellation method} \label{subsec:wc}

\begin{figure}
\centering
\includegraphics[scale=0.85]{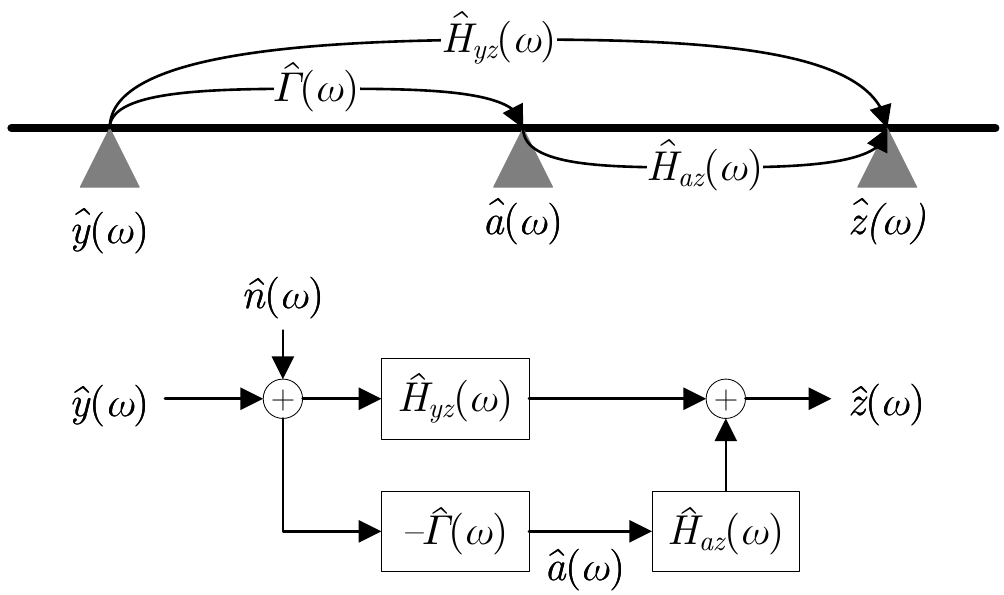}
\caption{Schematic representation of the control problem (top) and the corresponding flow chart (bottom).}
\label{fig:flowchart}

\end{figure}

The control problem is depicted via the schematic presented in figure \ref{fig:flowchart}, where $y$, $u$ and $z$ denote the sensor, actuator and observe locations and $n$ denotes the uncorrelated measurement noise. We adopt a reactive feed-forward loop similar to the one used in \citet{maia2021real}. The block diagram of the feed-forward loop is also given in figure \ref{fig:flowchart}. The relation between the sensor and the observer can be written as
\begin{align} \label{eq:ztot}
\hat{z}_c(\omega) = \big(\hat{H}_{yz}(\omega) + \hat{\Gamma}(\omega)\hat{H}_{az}(\omega)\big)\left(\hat{y}(\omega)+\hat{n}(\omega)\right),
\end{align}
where the hat denotes a Fourier transformed quantity, $\hat{z}_c$ denotes the observer when controller is active and $\hat{H}_{yz}$ and $\hat{H}_{az}$ denote the linear transfer functions between $y$-$z$, and $a$-$z$, respectively, which satisfy
\begin{align} \label{eq:zy}
\hat{z}_y(\omega)=\hat{H}_{yz}(\omega)\left(\hat{y}(\omega)+\hat{n}(\omega)\right), \\
\hat{z}_a(\omega)=\hat{H}_{az}(\omega)\hat{a}(\omega), \label{eq:zu}
\end{align}
where $\hat{z}_y$ and $\hat{z}_a$ denote the noise generated by the wavepacket and the actuator, respectively, when there is no loop connecting the measurement and actuation. In the case of stochastic forcing in \eqref{eq:linsys}, $\hat{y}(\omega)$ and $\hat{a}(\omega)$ becomes stochastic quantities as well. Calculating the cross-spectral densities (CSD) defined by $\hat{P}_{yy}\triangleq\langle\hat{y}\hat{y}^*\rangle$,  $\hat{P}_{nn}\triangleq\langle\hat{n}\hat{n}^*\rangle$, and $\hat{P}_{zy}\triangleq\langle\hat{z}_y\hat{y}^*\rangle$ for \eqref{eq:zy} and $\hat{P}_{aa}\triangleq\langle\hat{a}\hat{a}^*\rangle$ and $\hat{P}_{za}\triangleq\langle\hat{z}_a\hat{a}^*\rangle$ for \eqref{eq:zu}, the transfer functions can be calculated as
\begin{align}
\hat{H}_{yz}(\omega) = \hat{P}_{zy}(\omega)\left(\hat{P}_{yy}+\hat{P}_{nn}\right)^{-1}(\omega), \\
\hat{H}_{az}(\omega) = \hat{P}_{za}(\omega)\hat{P}_{aa}^{-1}(\omega),
\end{align}
where $\langle\cdot\rangle$ denotes the expectation operator and the superscript $*$ denotes complex conjugation. The CSD between two stochastic signal $\hat{u_1}(\omega)$ and $\hat{u_2}(\omega)$ is predicted using the Welch algorithm \citep{welch_ieee_1967}
\begin{align} \label{eq:welch}
\langle\hat{u}_1(\omega)\hat{u}_2^*(\omega)\rangle=\frac{1}{N_b}\sum_{k=1}^{N_b}\hat{u}_1^{(k)}(\omega){{}\hat{u}_2^*}^{(k)}(\omega),
\end{align}
where $N_b$ denotes the number of realisations. 

 Note that a feedback loop between the actuator and the measurement is not necessary to model the actual control problem. The piezoelectric actuator is expected to generate acoustic actuation without causing a hydrodynamic change on the wavepacket, and hence, on the sensor, which is meant to measure hydrodynamic fluctuations in the jet. When the actuation has no effect on the measurement, feedback control reduces to feed-forward control as discussed in \citet{sasaki2018closed}. Setting the output at the observer $\hat{z}$ as zero in \eqref{eq:ztot}, the control law associated with the feed-forward loop described above is given by
\begin{align} \label{eq:wckernel}
\hat{\Gamma}(\omega) = -\hat{H}_{yz}(\omega){\hat{H}_{az}^{-1}(\omega)}.
\end{align}

The control kernel in the wave cancellation method is obtained in the frequency domain. This implies that causality is not ensured. Depending on the positions of the sensor, observer and actuator, one can obtain a causal or non-causal control loop. A causal control can be implemented by computing the control kernel in the time domain by inverse Fourier transforming and then trimming the non-causal part of the kernel. The kernel however is no longer optimal once it is trimmed. We discuss the use of optimal causal control in the next subsection to mitigate this effect.
\subsection{Optimal causal control} \label{subsec:opcaus}
An optimal frequency-domain control approach where causality is imposed as a constraint was proposed in \citet{martini_jfm_2022}. Optimality is ensured via a resolvent-based estimation which was introduced earlier in another study \citep{martini_jfm_2020}. The method was extended to include causality in the form of a Lagrange multiplier which is computed via Wiener-Hopf decomposition. The method is briefly revisited here. For a complete description, we refer the reader to \citet{martini_jfm_2022}.

For modelling the wavepacket, we consider the linear system given in \eqref{eq:linsys}. Discretising in space and taking the Fourier transform, we obtain
\begin{align}
-i\omega\hat{\mathbf{q}}(\omega)-\mathbf{A}\hat{\mathbf{q}}(\omega)=\hat{\mathbf{f}}(\omega),
\end{align}
which can be reorganised as 
\begin{align}
(-i\omega\mathbf{I}-\mathbf{A})\hat{\mathbf{q}}(\omega)=\hat{\mathbf{f}}(\omega),
\end{align}
where here and in what follows bold letters denote the discretised vectors of the corresponding variables denoted with the same letter and $\mathbf{I}$ denotes the identity matrix. In what follows, the dependencies on $\omega$ will be dropped for brevity. Defining the resolvent operator as $\mathbf{R}=(-i\omega\mathbf{ I}-\mathbf{A})^{-1}$, one can obtain a linear relation between the external forcing and the state as
\begin{align}
\hat{\mathbf{q}}=\mathbf{R}\hat{\mathbf{f}}.
\end{align}
A sensor reading can be represented as 
\begin{align}
\hat{\mathbf{y}}=\mathbf{C}\hat{\mathbf{q}} + \hat{\mathbf{n}},
\end{align}
where $\mathbf{C}$ and $\hat{\mathbf{n}}$ denote the measurement matrix and the measurement noise, respectively. 

For the acoustic propagation problem, we consider a Green's function, $\hat{G}(x,x^\prime,\omega)$ calculating the acoustic pressure at the observer at $x^\prime$ generated by the source at $x$ and scattered by the semi-infinite plate. This relation is given in discrete form as
\begin{align}
\hat{\mathbf{z}}=\hat{\mathbf{G}}\hat{\mathbf{q}}.
\end{align}
Similarly, the effect of the actuation, $\hat{\mathbf{a}}$, on the observer is modelled with another Green's function,
\begin{align}
\hat{\mathbf{z}}_a=\hat{\mathbf{G}}_a\hat{\mathbf{a}}.
\end{align}
The details of the linear system and Green's functions will be given in section \ref{sec:imp}. Defining the control gain between the measurement and the actuation as
\begin{align}
\hat{\mathbf{a}}=\hat{\mathbf{\Gamma}}\hat{\mathbf{y}},
\end{align}
and the controlled observer as
\begin{align} \label{eq:zcdef}
\hat{\mathbf{z}}_c=\hat{\mathbf{z}}+\hat{\mathbf{z}}_a=\hat{\mathbf{z}}+\hat{\mathbf{G}}_a\hat{\mathbf{\Gamma}}\hat{\mathbf{y}},
\end{align}
the cost function for non-causal control, minimizing both the acoustic pressure at the observer and the actuation is given as
\begin{align}
\mathcal{J}=\int_{-\infty}^{\infty}\big\langle Tr\big(\hat{\mathbf{z}}_c^H\hat{\mathbf{z}}_c\big) + Tr\big(\hat{\mathbf{a}}^H\mathbf{Q}\hat{\mathbf{a}}\big)\big\rangle  d\omega,
\end{align}
where $\mathbf{Q}$ is the penalty for actuation and the superscript $H$ denotes the Hermitian transpose. Using the distributive property of expectation operator and the identity $Tr(\mathbf{E}\mathbf{F})=Tr(\mathbf{F}\mathbf{E})$ for any two matrices $\mathbf{E}$ and $\mathbf{F}$ of appropriate size, the cost function can be expanded as 
\begin{align} \label{eq:costexp}
\mathcal{J}=\int_{-\infty}^{\infty}\langle Tr(\hat{\mathbf{z}}\hat{\mathbf{z}}^H) + Tr(\hat{\mathbf{z}}_a^{ }\hat{\mathbf{z}}^H_a) + Tr(\hat{\mathbf{z}}\hat{\mathbf{z}}_a^H) + Tr(\hat{\mathbf{z}}_a\hat{\mathbf{z}}^H) + Tr(\mathbf{Q}\hat{\mathbf{a}}\hat{\mathbf{a}}^H)\rangle d\omega.
\end{align}
Defining the CSD matrices $\hat{\mathbf{P}}_{zz}\triangleq\langle\hat{\mathbf{z}}\hat{\mathbf{z}}^H\rangle$, $\hat{\mathbf{P}}_{yy}\triangleq\langle\hat{\mathbf{y}}\hat{\mathbf{y}}^H\rangle$ and $\hat{\mathbf{P}}_{zy}\triangleq\langle\hat{\mathbf{z}}\hat{\mathbf{y}}^H\rangle$, the terms appearing in \eqref{eq:costexp} can be written as
\begin{align} \label{eq:costst}
Tr(\langle\hat{\mathbf{z}}_a^{ }\hat{\mathbf{z}}^H_a\rangle) &=Tr(\hat{\mathbf{G}}_a^H\hat{\mathbf{G}}_a\hat{\mathbf{\Gamma}}\hat{\mathbf{P}}_{yy}\hat{\mathbf{\Gamma}}^H),\\
Tr(\langle\hat{\mathbf{z}}\hat{\mathbf{z}}^H_a\rangle) &=Tr(\hat{\mathbf{G}}_a^H\hat{\mathbf{P}}_{zy}\hat{\mathbf{\Gamma}}^H),\\
Tr(\mathbf{Q}\langle\hat{\mathbf{a}}\hat{\mathbf{a}}^H\rangle) &=Tr(\mathbf{Q}\hat{\mathbf{\Gamma}}\hat{\mathbf{P}}_{yy}\hat{\mathbf{\Gamma}}^H).
\end{align}
The cost function $\mathcal{J}$ can be minimized by treating $\hat{\mathbf{\Gamma}}$ and $\hat{\mathbf{\Gamma}}^H$ as independent variables \citep{ahlfors1979} and setting the derivative of \eqref{eq:costexp} with respect to $\hat{\mathbf{\Gamma}}^H$ to zero. Substituting This leads to 
\begin{align}
(\hat{\mathbf{G}}_a^H\hat{\mathbf{G}}_a^{ }+\mathbf{Q})\hat{\mathbf{\Gamma}}\hat{\mathbf{P}}_{yy} + \hat{\mathbf{G}}_a^H\hat{\mathbf{P}}_{zy}=0,
\end{align}
which leads to the optimal gain matrix given as
\begin{align} \label{eq:knc}
\hat{\mathbf{\Gamma}}= - (\hat{\mathbf{G}}_a^H\hat{\mathbf{G}}_a^{ }+\mathbf{Q})^{-1}\hat{\mathbf{G}}_a^H\hat{\mathbf{P}}_{zy}^{ }\hat{\mathbf{P}}_{yy}^{-1}.
\end{align}

Up to this point, no causality constraint is imposed while deriving \eqref{eq:knc}, and therefore, the resulting control matrix can contain a non-causal part depending on the positions of the sensors, actuators and observers. A causal control designed in the frequency domain is achieved by constructing a Wiener-Hopf problem. The cost function in this case is given as 
\begin{align} \label{eq:costup}
\mathcal{J}^\prime=\mathcal{J} + \int_{-\infty}^{\infty}Tr(\hat{\mathbf{\Lambda}}_-^{ }\hat{\mathbf{\Gamma}}_+^{ } + \hat{\mathbf{\Gamma}}_+^H\hat{\mathbf{\Lambda}}_-^H)d\omega,
\end{align}
where $\hat{\mathbf{\Gamma}}_+$ denotes the causal gain matrix and $\hat{\mathbf{\Lambda}}_-$ is a matrix required for the problem to be well-posed in the entire complex plane (see \citet{martini_jfm_2022} for details). The updated cost function given in \eqref{eq:costup} leads to the final equation 
\begin{align} \label{eq:wh}
(\hat{\mathbf{G}}_a^H\hat{\mathbf{G}}_a^{ }+\mathbf{Q})\hat{\mathbf{\Gamma}}_+\hat{\mathbf{P}}_{yy} + \hat{\mathbf{\Lambda}}_- + \hat{\mathbf{G}}_a^H\hat{\mathbf{P}}_{zy}^{ }=0
\end{align}
to be solved for $\hat{\mathbf{\Gamma}}_+$. The methodology proposed by \citet{martini_jfm_2022} to solve the Wiener-Hopf problem in \eqref{eq:wh} is summarized in Appendix \ref{secA1}. The resulting control kernel is given as
\begin{align} \label{eq:ockernel}
\hat{\mathbf{\Gamma}}_+=\hat{\mathbf{D}}_+^{-1}\big(\hat{\mathbf{D}}_-^{-1}\hat{\mathbf{F}}\hat{\mathbf{E}}_-^{-1}\big)_+\hat{\mathbf{E}}_+^{-1},
\end{align}
where $\hat{\mathbf{D}}\triangleq\hat{\mathbf{G}}_a^H\hat{\mathbf{G}}_a^{ }+\mathbf{Q}$, $\hat{\mathbf{E}}\triangleq\hat{\mathbf{P}}_{yy}$ and $\hat{\mathbf{F}}\triangleq-\hat{\mathbf{G}}_a^H\hat{\mathbf{P}}_{zy}^{ }$.

\section{Linear model for installation noise} \label{sec:linmod}
\begin{figure}
\centering
\includegraphics[scale=0.60]{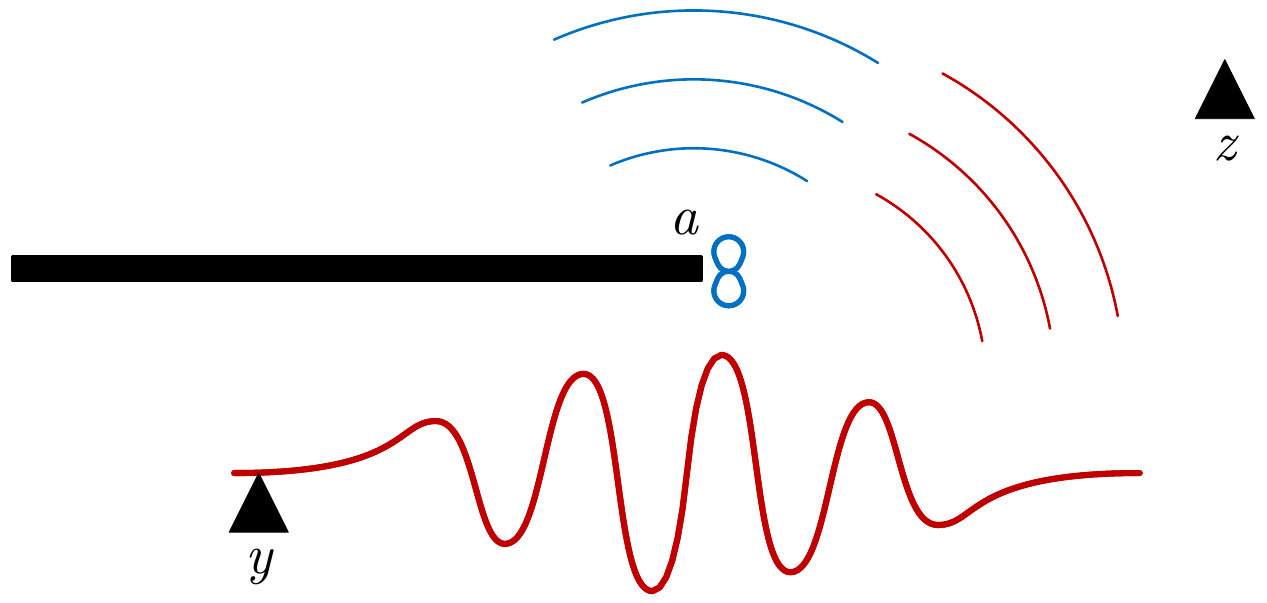}
\caption{Schematic of the model problem that consists of a wavepacket, semi-infinite plate and a point dipole with the sensor $y$ and the observer $z$ is positioned on the wavepacket and in the acoustic field, respectively.}
\label{fig:model}
\end{figure}

We model in this study the installation noise in two parts: The wavepackets that populate the jet turbulence and interact with flaps yielding a dominant noise source are modelled using the linear Ginzburg-Landau (G-L) equation forced with external stochastic forcing. The G-L equation in linear form constitutes a linear advection-diffusion-dissipation system. A similar methodology was used in the by \citet{evert_aiaa_2016} for source modelling.  The acoustic interaction between the wavepacket and the flap is modelled by placing a semi-infinite half plane near the wavepacket and using a tailored Green's function which accounts for the scattering due to the plane. A schematic is illustrated in figure \ref{fig:model}. The details of each part are given in the following subsections. 

\subsection{Ginzburg-Landau model} \label{subsec:gl}
The linear G-L equation is given as,
\begin{align} \label{eq:gl}
\partial_t q + U\partial_x q - \gamma\partial_{xx}q -\mu q =  f,
\end{align}
where $U=1$ is the convection coefficient, $\gamma=1/120$ denotes the viscosity-like parameter, $\mu=\mu_0(1-2x/L)$ with $L=20$ determines the growth of the response over the domain $x=[0,L]$, and $f$ is the external forcing term. The global stability of the homogeneous part is determined by $\mu_0$. The critical limit for stability is reached at $\mu_0=2.29$. For values of $\mu_0$ less than the critical limit, the G-L equation behaves like an amplifier flow, while for values higher than the critical limit, it represents an oscillatory system. As jets are of amplifier type, we set $\mu_0=1.5$ in this study to obtain a convectively unstable system. The boundary conditions at $x=0$ and $x=L$ are set to 0. The domain is discretised using Chebyshev discretisation with the number of grid points, $N=100$. The implicit Crank-Nicholson method was used for time marching with a time step $\Delta t=0.125$. The fast Fourier transform (FFT) was used with 1024 FFT points when transforming the time-domain data into the frequency domain. For taking the Fourier transform (FT) of stochastic data, we used an exponential windowing function \citep{martini_arxiv_2019} given as
\begin{align}
w=e^{4-\frac{T^2}{t(T-t)}},
\end{align}
where $t$ and $T$ denote the time and the period, respectively.

Equation \eqref{eq:gl} can be rewritten in discretised form as
\begin{align} \label{eq:gllindis}
\partial_t\mathbf{q}+\mathbf{A}\mathbf{q}=\mathbf{f},
\end{align}
where $\mathbf{f}$ and $\mathbf{q}$ denote the discretised forcing and state, respectively, and $\mathbf{A}$ denotes the discretised linear operator. 

\begin{figure}
\centering
\includegraphics[scale=0.85]{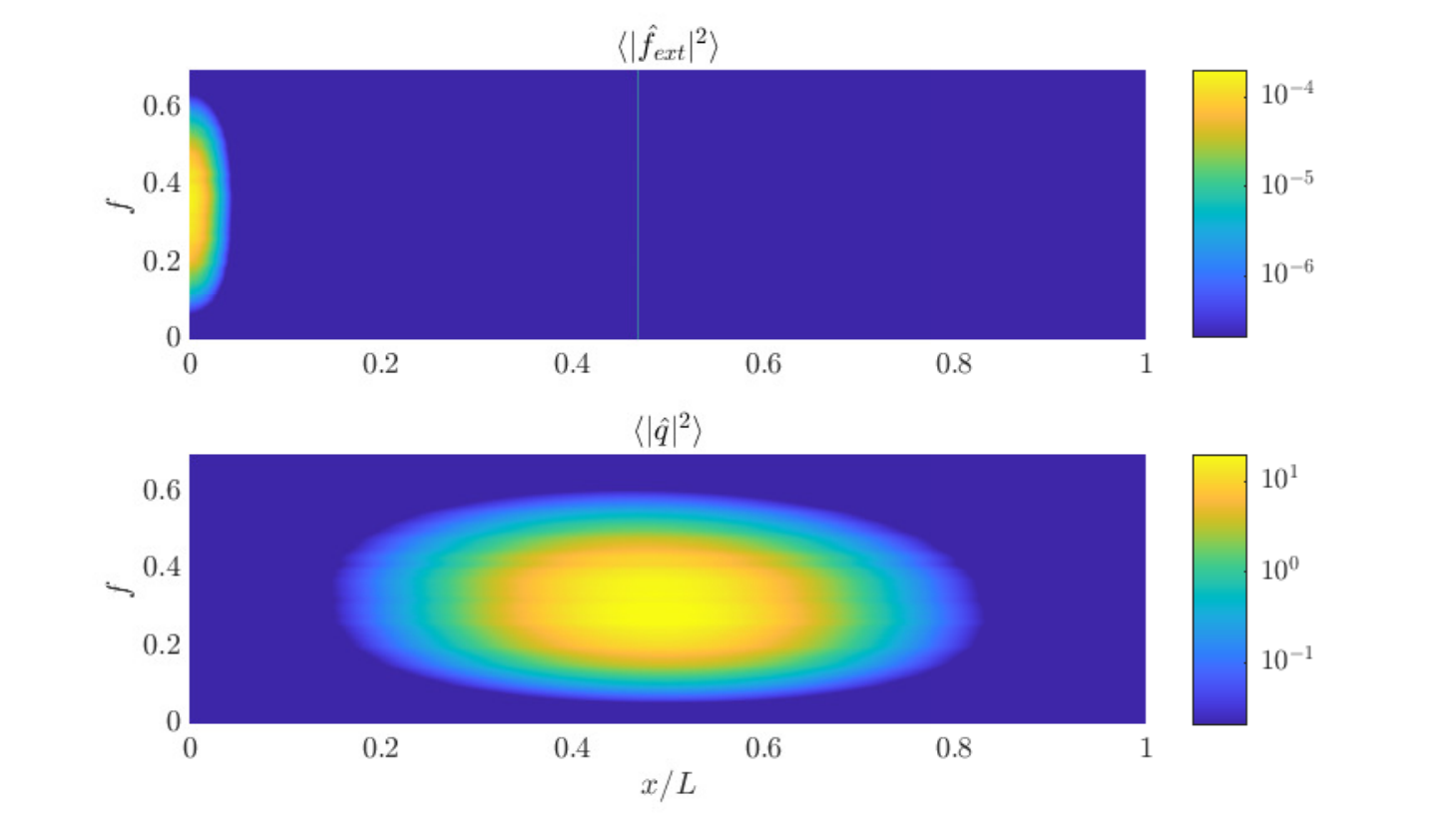}
\caption{PSD map of the external forcing $f_{ext}$ (top) and state $q$ (bottom).}
\label{fig:psdmap}
\end{figure}

The external forcing was generated as a random signal which was then band-pass filtered in time to limit the frequency content to the interval $0.1<f<0.5$. This frequency range was selected since the sound generated by the wavepacket-semi-infinite plate configuration is of dipolar type in this range as will be seen in section \ref{subsec:tgf}, similar to the actual case of a jet-flap-interaction noise \citep{piantanida2016scattering}.  The spatial support and correlation are controlled via multiplication by a two-point correlation tensor given as
\begin{align}
\mathcal{M}(x_1,x_2)=e^{-(x_1^2+x_2^2)/L_x^2}e^{-(x_1-x_2)^2/L_c^2},
\end{align}
where $L_x=L/40$ and $L_c=L/10$ denote the spatial support and correlation lengths, respectively. Discretising $\mathcal{M}$ and the filtered forcing, the external forcing used in \eqref{eq:gllindis} is given as 
\begin{align}
\mathbf{f}=\mathbf{M}\tilde{\mathbf{f}},
\end{align}
where $\mathbf{M}$ denotes the discretised correlation tensor and $\tilde{\mathbf{f}}$ denotes filtered-in-time white-in-space forcing described above. The power spectral densities (PSD) of the external forcing and the state are shown in figure \ref{fig:psdmap}. The forcing is limited to the initial part of the domain within $x/L<1$ and is limited to the frequency range $0.1<St<0.6$ outside of which is filtered. Here, and in what follows, we present frequencies in terms of Strouhal number, $St\triangleq fU/D$, where $f$ denotes the frequency, and $U$ and $D$ denote the characteristic velocity and length, respectively, where both are set as unity. The forcing PSD peaks around $St=0.3$. The disturbances caused by this external forcing in the inlet section are amplified due to the convective instability of the G-L system and peak around the mid-domain, where $\mu$ becomes zero. Beyond this point, $\mu<0$ causes damping of the perturbations before reaching the end of the domain. The PSD distribution of the state in the frequency axis is similar to that of the forcing, limited to the range $0.1<St<0.6$ and peaking around $0.3$.

\subsection{Tailored Green's function for semi-infinite plate} \label{subsec:tgf}
The tailored Green's function (TGF) to compute the sound generated by a point source positioned near a semi-infinite plate was introduced by \citet{fwilliams_jfm_1970} and is given as,
\begin{align} \label{eq:tgf}
G_t(\mathbf{x},\mathbf{y},\omega)=\frac{e^{\frac{i\pi}{4}}}{\sqrt{\pi}}\left(\frac{e^{-ikR}}{R}\int_{-\infty}^{u_R}e^{-iu^2}du + \frac{e^{-ikR^\prime}}{R^\prime}\int_{-\infty}^{u_{R^\prime}}e^{-iu^2}du\right),
\end{align}
where
\begin{align} \label{eq:ur}
u_R&=2\sqrt{\frac{krr_0}{B+R}}\cos\frac{\theta-\theta_0}{2}, \\
u_{R^\prime}&=2\sqrt{\frac{krr_0}{B+R^\prime}}\cos\frac{\theta+\theta_0}{2}. \label{eq:urp}
\end{align}

\begin{figure}
\centering
\includegraphics[scale=0.5]{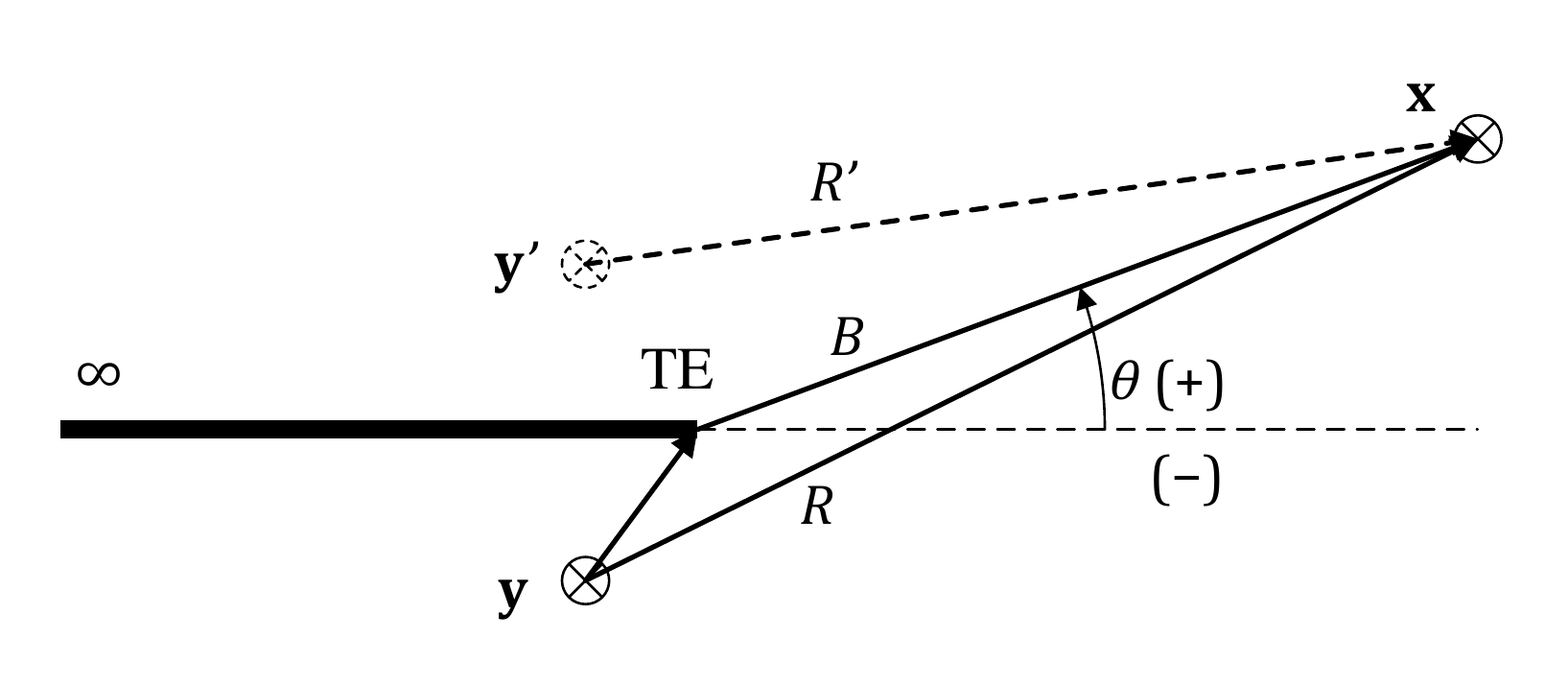}
\caption{Schematic to illustrate the geometric definitions related to the tailored Green's function.}
\label{fig:tgfschem}
\end{figure}
In the above equations, $\mathbf{y}=[r_0,\theta_0,z_0]^\top$ and $\mathbf{x}=[r,\theta,z]^\top$ denote the source and observer positions, respectively, $\omega$ is the angular frequency, and $k=\omega c_0$ is the acoustic wavenumber, where the speed of sound, $c_0$ is set as 2.5 for this study yielding a Mach number of $M\triangleq U/c_0=0.4$. $R$ and $R^\prime$ denote the distance between the source and the observer, and between the source image and the observer, respectively. $B$ denotes the shortest distance between the source and the observer passing through the trailing edge. A schematic is provided in figure \ref{fig:tgfschem} to illustrate the corresponding geometry and the axis definitions. With the centre of the cylindrical coordinate system located at the TE of the semi-infinite plate, the terms $R$, $R^\prime$ and $B$ can be calculated as,
\begin{align}
R &= \sqrt{r^2+r_0^2-2rr_0\cos(\theta-\theta_0)+ (z-z_0)^2}, \\
R^\prime &= \sqrt{r^2+r_0^2-2rr_0\cos(\theta+\theta_0)+ (z-z_0)^2}, \\
B &= \sqrt{(r+r_0)^2+ (z-z_0)^2}.
\end{align}
The Fresnel integrals in \eqref{eq:tgf} are computed using a series expansion as in \citet{chang_1996}. Note that in case the distance between the source and the semi-infinite plate tends to infinity, the Green's function given in \eqref{eq:tgf} reduces to the free-field Green's function as
\begin{align}
G(\mathbf{x},\mathbf{y},\omega)=\frac{e^{-ikR}}{R}.
\end{align}
The acoustic field generated by the line source obtained from the G-L model can be calculated via the following integration,
\begin{align} \label{eq:pint}
\hat{p}(\mathbf{x},\omega)=\int_S \hat{q}(\mathbf{y},\omega)G_t(\mathbf{x},\mathbf{y},\omega)d\mathbf{y},
\end{align}
where $S$ denotes the domain where the line source is computed. The line source in the present case is stochastic. To visualise the expected acoustic field, we perform spectral proper orthogonal decomposition (SPOD) \citep{towne_jfm_2018} and calculate the optimal  SPOD mode of the line source. We then compute the integral given in \eqref{eq:pint}. Details about how to perform SPOD using the discrete system are given in Appendix \ref{secA2}.
 
\begin{figure}
\centering
\includegraphics[scale=0.80]{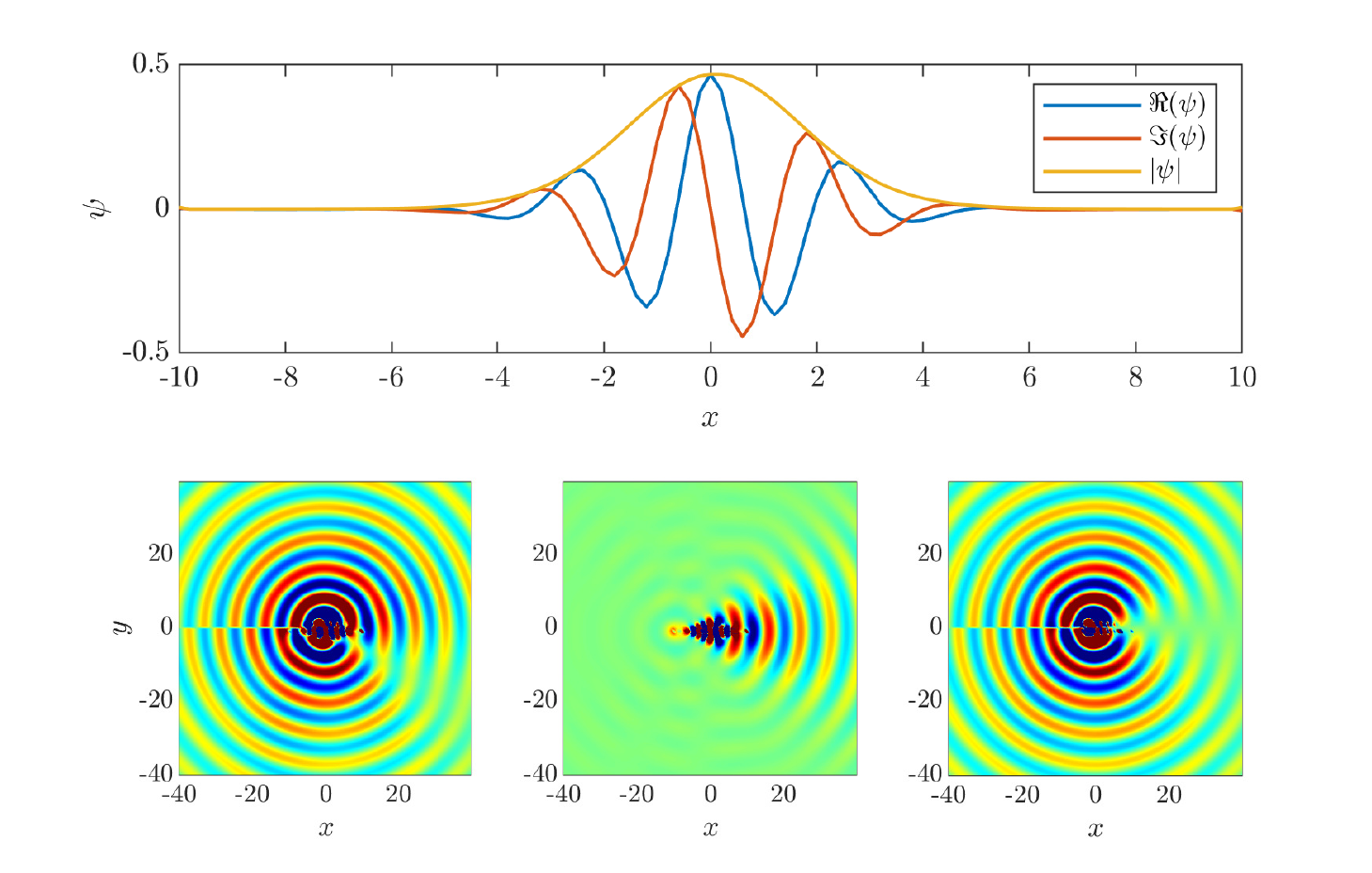}
\caption{Top: The optimal SPOD mode $\psi$ of the state $q$ at $St=0.3$. Bottom: total (left), incident (centre) and scattered (right) pressure fields obtained using the line source in the top plot positioned near a semi-infinite plane.}
\label{fig:tgfpfield}
\end{figure}

Similar to the case investigated in \citet{nogueira_jsv_2017}, the relative position in Cartesian coordinates of the source with respect to TE of the semi-infinite plane is given as $(x_0,y_0)=([-10,10],-1)$ corresponding to a source domain of length 20 as denoted in section \ref{subsec:gl}. The resulting acoustic pressure for the total, incident and scattered fields are shown in figure \ref{fig:tgfpfield}. Note that the TGF is only valid for the far-field pressure. Therefore the pressure field close to the origin where the TE is located should be disregarded. It is seen that the scattered field is anti-symmetric in the $y$-direction and is similar to a dipole, particularly in the downstream region. This provides a supporting argument for our control approach as we try to cancel this pressure field with a dipole located at the TE of the semi-infinite plane. 

\begin{figure}
\centering
\includegraphics[scale=0.80]{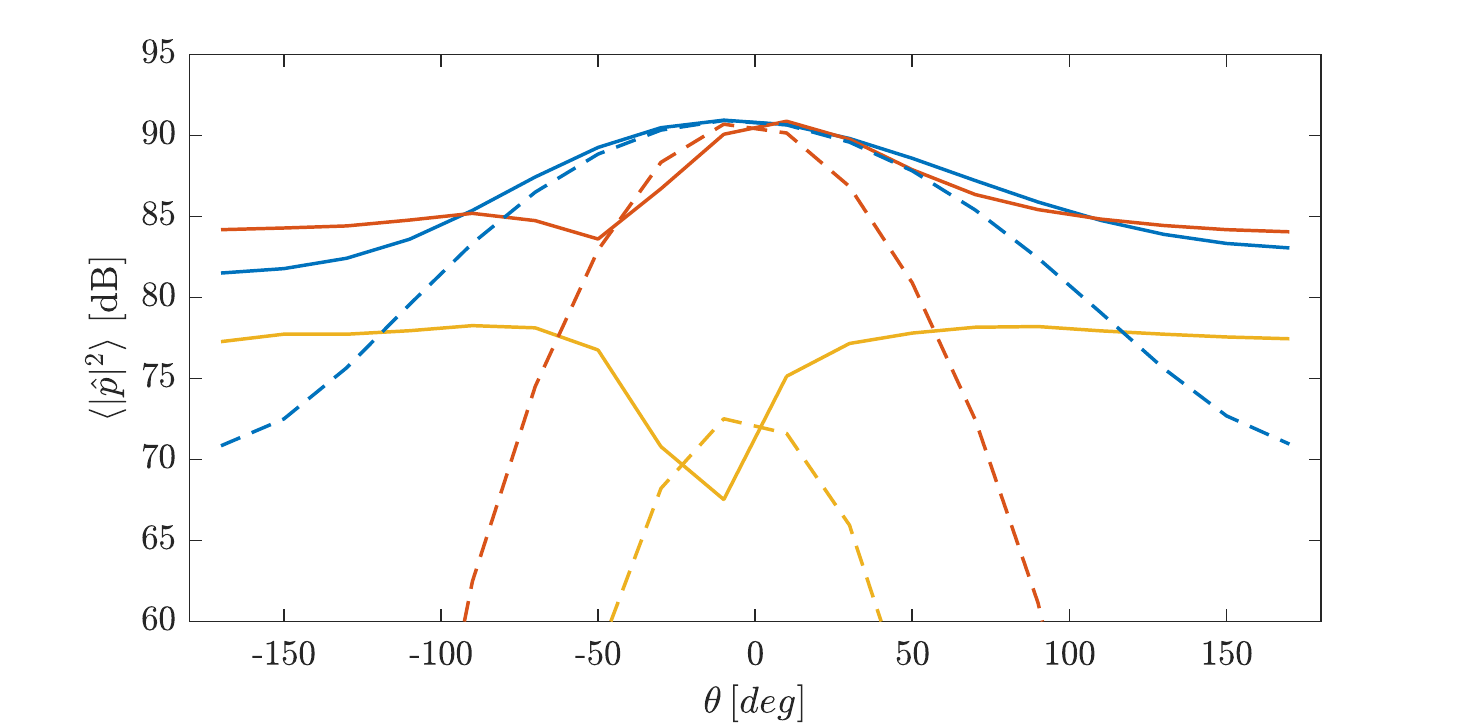}
\caption{PSD of the acoustic field due to the line source with (solid) or without (dashed) the semi-infinite plane at three frequencies: $St=0.1$ (blue), 0.2 (orange), 0.3 (yellow).}
\label{fig:glacous}
\end{figure}

The directivity of the acoustic field with or without the semi-infinite plane is shown in figure \ref{fig:glacous} at a number of frequencies. The directivity angle, $\theta$ is defined as shown in figure \ref{fig:tgfschem}. The line source itself has a strong directivity with peak radiation in the aft-angle region reminiscent of subsonic jets \citep{cavalieri_jfm_2012}. Installing the plane significantly enhances the sound field in the upstream region (large $\lvert\theta\rvert$) while its effect is negligible in the downstream regions (small $\lvert\theta\rvert$). This is expected as the infinite plane does not scatter any sound field in the $\theta=0$ direction.

\subsection{Green's function for the actuator} \label{subsec:dgf}
To obtain a transfer function between the actuator and the observer, we need another Green's function. The directivity of a dipole aligned with the surface normal is given by $\sin \theta$, where $\theta$ is defined according to the coordinate axis shown in figure \ref{fig:tgfschem}. Given a time-variant function $\Gamma(t)$ to define the oscillation amplitude of the actuator, the acoustic pressure at the observer can be calculated by 
\begin{align}
p(\mathbf{x},t)=\frac{1}{rc_0}\Gamma\left(t-\frac{r}{c_0}\right)\sin\theta,
\end{align}
which is the retarded Green's function for a point dipole located at the TE of the surface. Note that there exists no scattering due to the plate itself as the dipole does not emit sound along the surface direction $\theta=\pi$. 

\section{Control methods applied to model problem} \label{sec:imp}
We present the details of the implementation of the control approaches described in \cref{sec2} to the model problem. We choose the test cases to highlight the importance of imposing causality while calculating the control kernel, as in optimal causal control, for improved control performance in the actual installation noise problem.

A schematic was given in figure \ref{fig:model}. The coordinate axis is the same as the one shown in figure \ref{fig:tgfschem}, where the semi-infinite plate lies on the $x$-axis with its TE at $x=0$. The line source is placed at $(x_0,y_0)=([-10,10],-1)$. For all the cases, we fix the observer position to $(r,\theta)=(20,-\pi/2)$. Note that $\theta=-\pi/2$ indicates the direction of the ground, which is the relevant direction for the installation noise problem. The position of the sensor directly affects control performance as it determines whether the control problem is causal or not. A detailed analysis is given below.

\subsection{Causality of the control problem}
As described earlier, we consider a reactive feedforward control: the sensor measures the fluctuations in the wavepacket. The actuator then uses this information to reduce the noise of the wavepacket by generating an anti-sound. Placing the sensor upstream of the wavepacket leaves time for the actuator to generate the anti-sound to cancel the wavepacket noise. When the sensor is placed downstream of the beginning of the wavepacket, on the other hand, by the time of the sensor measurement, the wavepacket already generates some noise, which implies non-causal information is necessary in order to fully cancel this noise. 

\begin{figure}
\centering
\includegraphics[scale=0.80]{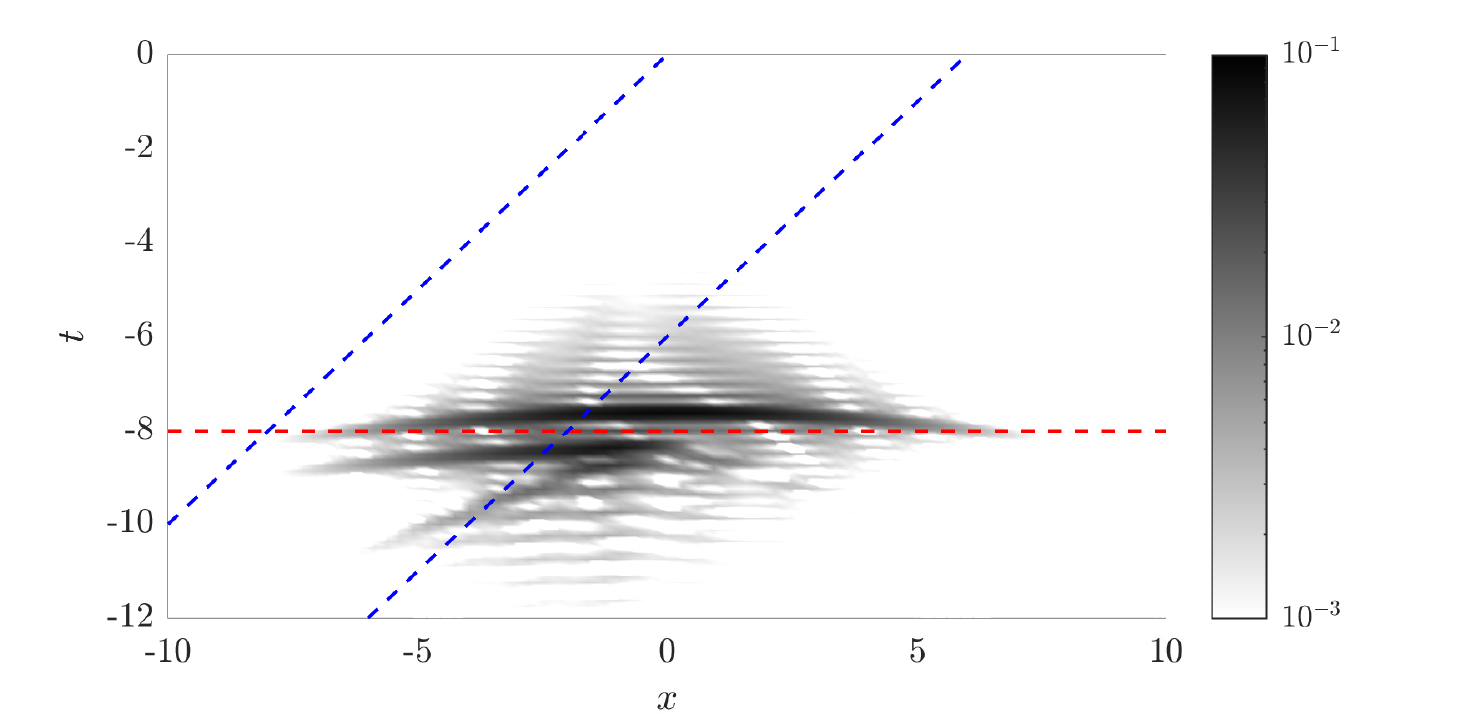}
\caption{Kernel of the TGF multiplied with the RMS of the line source for an observer at $(r,\theta)=(20,-\pi/2$. The red dashed line indicates the time delay between the actuator located at the TE and the observer. The blue dashed lines indicate convective time delay between the sensor and the source downstream of it.}
\label{fig:tgfkernel}
\end{figure}

This issue is visualised in figure \ref{fig:tgfkernel} where the time kernel of the tailored Green's function given in \eqref{eq:tgf} multiplied by the root-mean-square (RMS) of $\mathbf{q}(t)$ is shown as a function of position on the source domain. The map shows all the information in the past, required for calculating the sound at the observer at $t=0$. The kernel has two dominant peaks up to $x=10$, which corresponds to the line source and its image due to the presence of the semi-infinite plate. Beyond $x=10$, the image source disappears since the plate ends at this point. We observe an increasing delay in the peak location towards both ends of the domain. This is because the observer is located at $x=10$, and thus the acoustic waves generated at the borders of the domain have to travel a longer distance to reach the observer compared to those generated at the centre of the domain. The red dashed line indicates the time delay in the control action, given by the distance between the actuator and the observer, divided by the speed of sound. Each blue dashed line corresponds to a different source position, which is indicated by the intersection point of the line with the red dashed line. The slope of the blue dashed lines indicates the inverse of the convection velocity of the wavepacket. For a given sensor position, the part of the kernel below the corresponding blue dashed line is causal, and hence, can be predicted using the past information from the sensor readings. Contrary to that, the part of the kernel that is above the blue dashed line indicates future information, and therefore, is not available. We see that placing the sensor to $x=-8$ yields a system that is causal, while a significant portion of the kernel of the TGF remains on the non-causal side when the sensor is placed at $x=-2$. The latter position is more relevant regarding the actual installation noise problem since it approximately corresponds to the nozzle exit (see the discussion in the introduction). The sensor can be placed at the nozzle exit with minimum structural complexity, and one may expect a decent characterisation of the wavepacket using the velocity data at this position \citep{cavalieri_jfm_2013,nogueira_aiaa_2016}. Also the coherence, and thus the transfer function, between the sensor and the observer improves as the sensor is placed further downstream again exacerbating the causality problem. All these issues suggest that it is inevitable to suffer from causality constraint in the actual problem. In the following, we will investigate how the two control approaches perform in this configuration.

\subsection{Identifying the transfer functions}
The wave cancellation method requires identification of transfer functions between the sensor and the observer, $H_{yz}$, and similarly the actuator and the observer, $H_{uz}$. The latter is also required in the optimal causal control as discussed in \cref{subsec:opcaus}, while the former is modelled using the governing equation of the G-L system. 

Identification of the transfer functions requires computing the CSD matrices as discussed in section \ref{subsec:wc}. The CSD matrices are predicted using the Welch algorithm, which involves chopping a long time-domain signal into shorter blocks, taking the FT of each block, and averaging the resulting Fourier realisations. In case there exists a delayed correlation between the sensor and the observer, or similarly the actuator and the observer, one can account for this by choosing the length of the time block much larger than the time delay such that the effect of the time delay on the correlation level becomes negligible. For stochastic processes, given the length of the time block, $T$ and the time delay, $\tau$, the convergence of the correlation level between two signals is proportional to $T/\tau$. This solution can be feasible for experimental studies, while for numerical studies, increasing the length of the time block to reach a convergence at the correlation level can significantly increase the simulation cost. It also renders the overall cost a function of the observer position, which is not preferable. 

\begin{figure}
\centering
\includegraphics[scale=0.80]{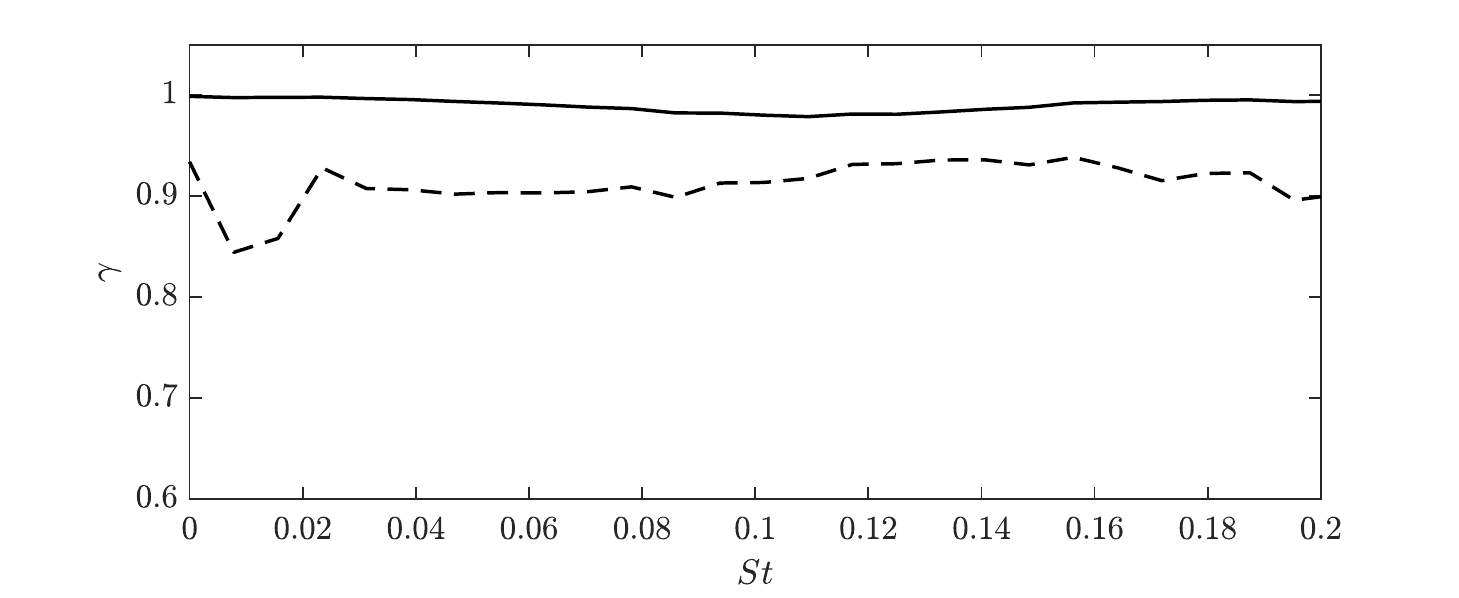}
\caption{The coherence value $\gamma_{yz}$ with (solid) and without (dashed) the phase correction.}
\label{fig:cohlevel}
\end{figure}

Alternatively, given two signals $a(t)$ and $b(t)$ with maximal correlation achieved between $a(t)$ and $b(t-\tau)$, the converged CSD matrix can be computed by taking the FTs
\begin{align}
\hat{a}&=\mathcal{F}(a(t)), \\
\hat{b}&=e^{i\omega\tau}\mathcal{F}(b(t-\tau)), \label{eq:tdcorr}
\end{align}
and using \eqref{eq:welch}. The exponential term in \eqref{eq:tdcorr} is added to correct the phase of the FT of the retarded signal $b(t-\tau)$. With this approach, a converged transfer function can be computed with smaller time blocks in the Welch averaging \citep{blanco_jfm_2022}, allowing better statistical convergence. In figure \ref{fig:cohlevel}, we show the effect of this correction on the coherence level defined as
\begin{align}
\gamma_{yz}=\hat{P}_{yz}(\hat{P}_{yy}\hat{P}_{zz})^{-1/2}.
\end{align}
The coherence level is an indicator of the linearity between two signals, where a unit coherence signifies a linear relation and zero coherence signifies no linearity. Since we use a linear model for the source, the expected coherence is 1, which is seen to be nearly the case when the phase correction described above is implemented. 

\begin{figure}
\centering
\includegraphics[scale=0.80]{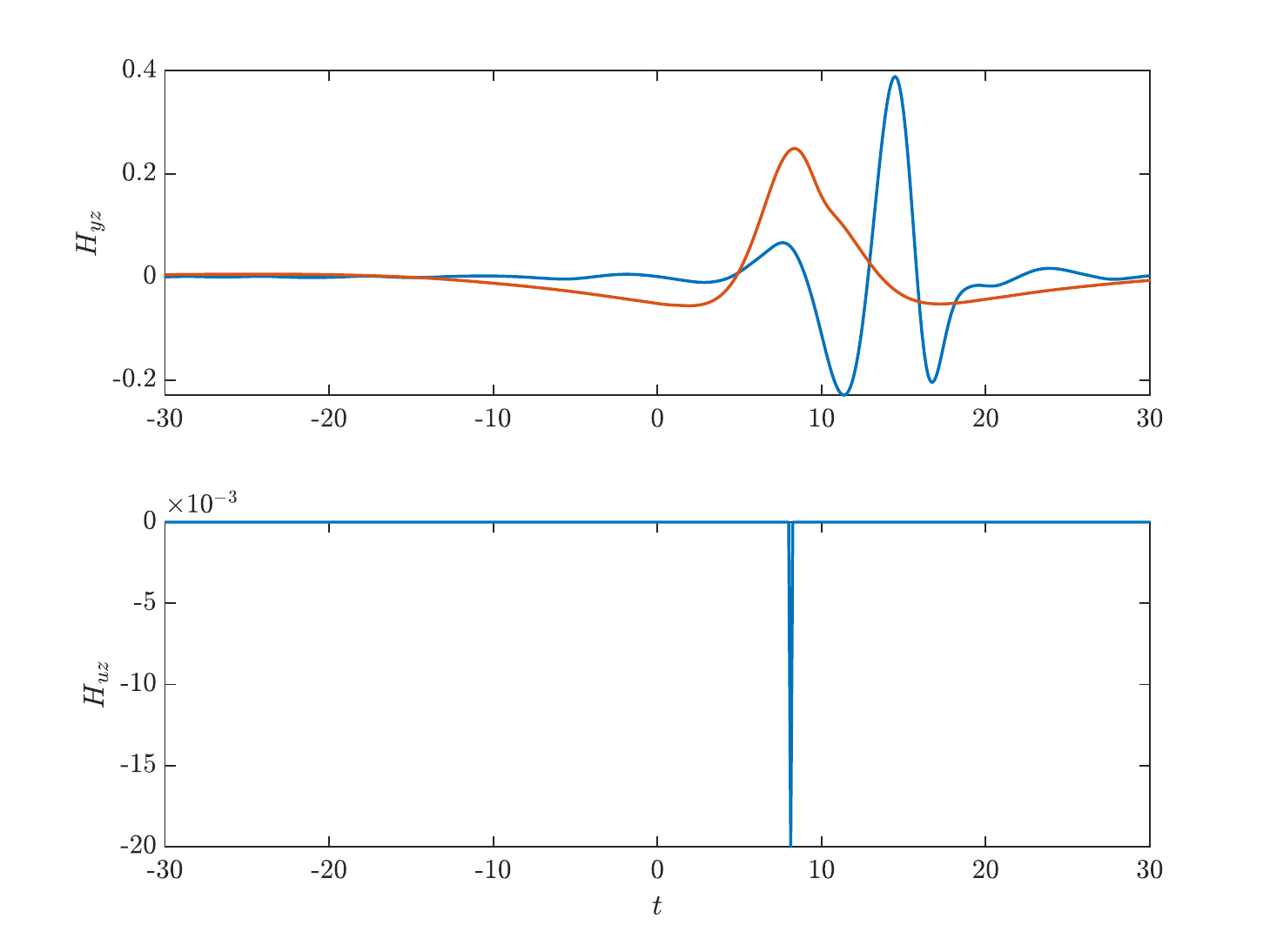}
\caption{The kernel $H_{yz}$ (top) for sensor positions at $x=-8$ (blue) and $-2$ (orange) and $H_{uz}$ (bottom). $H_{yz}$ at $x=-2$ is multiplied by 10 for better visual comparison.}
\label{fig:tfunc}
\end{figure}

The resulting time-domain transfer functions $H_{yz}$ for two sensor positions at $x=-8$ and $-2$ and $H_{uz}$ are shown in figure \ref{fig:tfunc}. The negative and positive parts of the time axis show the future and the past, i.e., the non-causal and causal parts, respectively. The transfer function between the actuator and the observer is a delta function with time delay, which is consistent with the definition given in section \ref{subsec:dgf}. The time instant of this peak defines the causality threshold for the actuator: any event before the peak cannot be canceled by the actuator. The transfer function between the sensor and the observer is seen to peak at a later time when the sensor is at $x=-8$, while the peak shifts to an earlier time when the sensor is moved to $-2$, roughly corresponding to the peak in $H_{uz}$. 

\subsection{Control performance}

\begin{figure}
\centering
\includegraphics[scale=0.80]{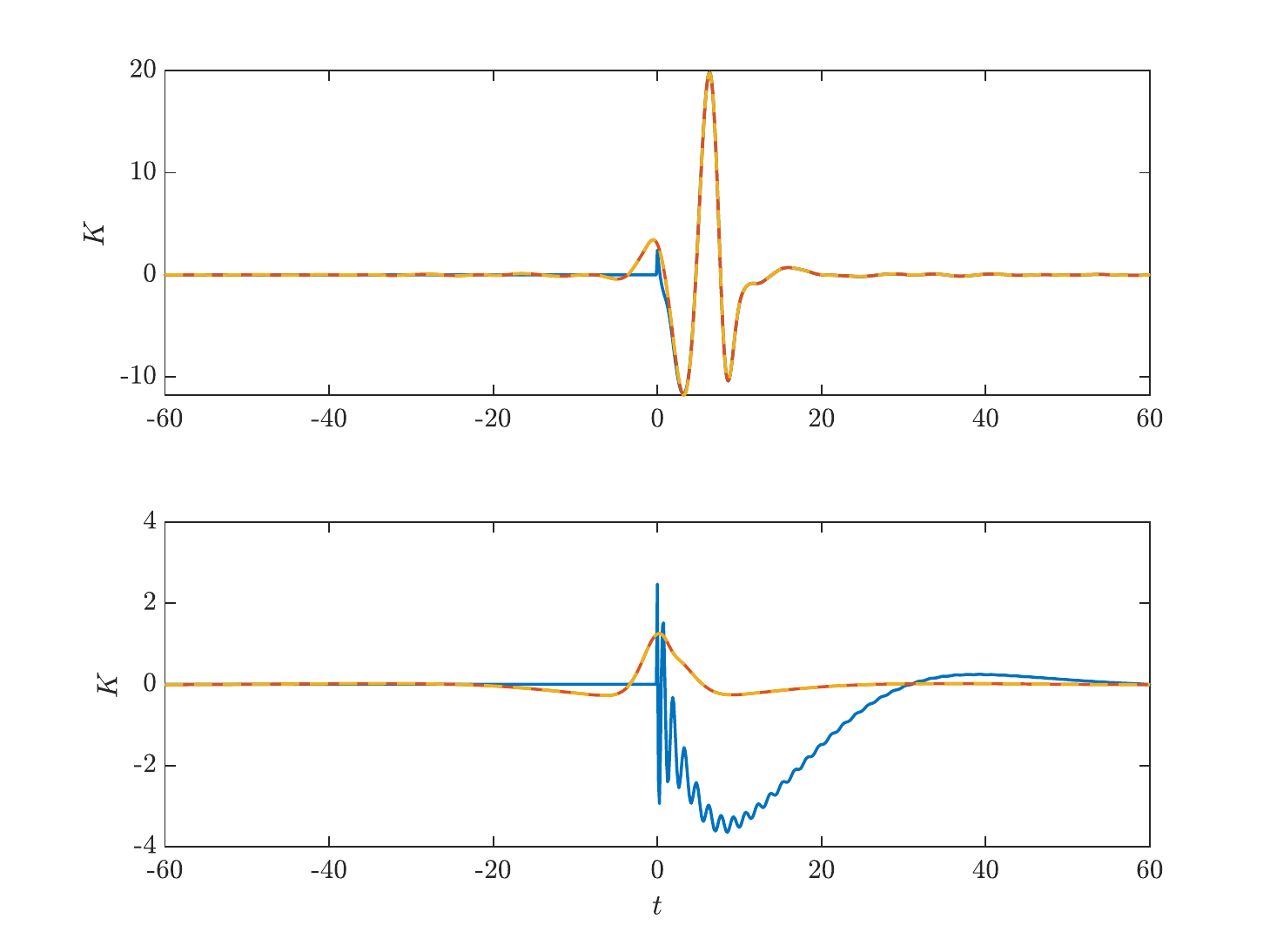}
\caption{Kernels for the wave-cancellation (yellow dashed), non-causal (orange) and optimal causal (blue) control with sensor positions at $x=-8$ (top) and $-2$ (bottom).}
\label{fig:kernels}
\end{figure}

The control kernels calculated using equations \eqref{eq:wckernel}, \eqref{eq:knc} and \eqref{eq:ockernel}, which we will hereafter refer to as wave-cancellation, non-causal and optimal causal, respectively, for the two sensor positions are compared in figure \ref{fig:kernels}. Once again, the negative and positive halves of the time axis denote the non-causal and causal parts, respectively. We see that the non-causal kernels given by \eqref{eq:wckernel} and \eqref{eq:knc} are identical at both positions, and hence, regardless of the causality of the configuration. When the sensor is positioned at $x=-8$, the optimal causal kernel given by \eqref{eq:ockernel} mainly has the same structure as the two other kernels except for a small peak at $t=0$.  Note that the non-causal kernel almost entirely lies in the causal part except for a small part contained in the non-causal half. This is thanks to the almost causal nature of the problem at this sensor position, as discussed earlier in this section. The peak in the causal kernel at $t=0$ is to compensate for the small bit of information that lies in the non-causal half. The existence of such sharp peaks in the time-domain kernel implies an increase in the energy of the high-frequency content of the kernel, which might be an issue in an actual actuator, which would have nonzero damping inertia to limit response beyond a certain frequency. Such a limitation, on the other hand, can be accounted for by defining a frequency-dependent actuation penalty, $\mathbf{Q}$. We set a constant actuation penalty of $\mathbf{Q}=\mathbf{I}$ in all the cases investigated in this study, as the aim is to find the maximum control performance that can be achieved in the case of an ideal actuator. 

\begin{figure}
\centering
\includegraphics[scale=0.80]{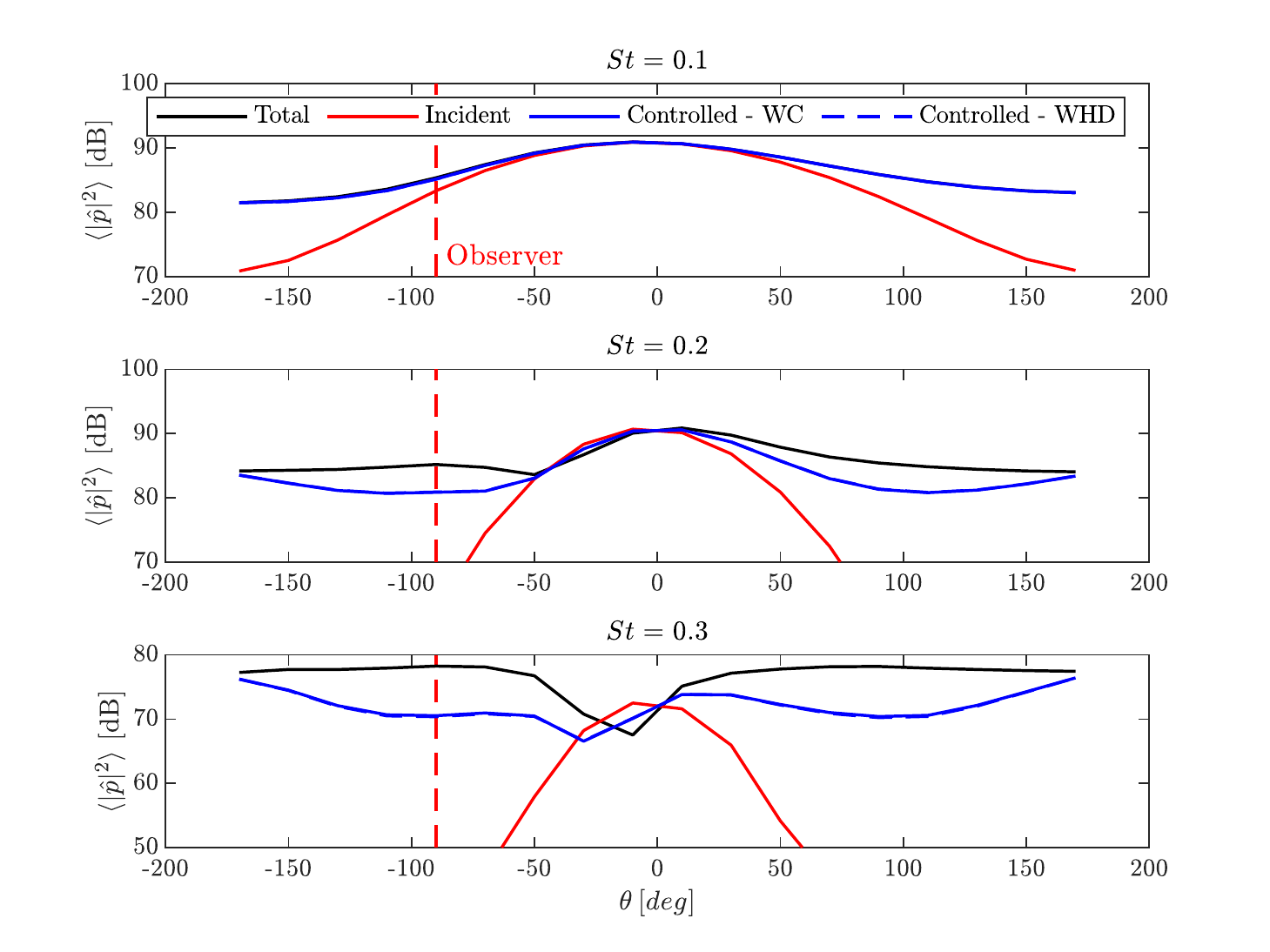}
\caption{Acoustic directivity for installed case with control compared against the installed (black) and free (red) cases without control at $St=0.1$ (top), 0.2 (middle) and 0.3 (bottom) when the sensor is positioned at $x=-8$. Blue solid and blue dashed lines correspond to the WC and OCC methods, respectively. The vertical red dashed line indicates the observer position.}
\label{fig:result2}
\end{figure}

The result of the noise control at $(r,\theta)=(20,-\pi/2)$ with sensor position at $x=-8$ is shown at three frequencies $St=0.1$, 0.2 and 0.3 in figure \ref{fig:result2} for both control methods. We set the measurement noise $\hat{\mathbf{P}}_{nn}=c\mathbf{I}$ with $c=1\times10^{-3}$ which applies to both methods. A discussion about the effect of these parameters on the control performance will be provided later. When calculating the control signal, the non-causal part of the kernel is truncated to zero to ensure causality in the real-time control. For the sensor positioned at $x=-8$, both methods yield a similar performance at all three frequencies. The vertical dashed line indicates the observer position. Noise reduction is negligible at $St=0.1$ while it increases up to 8 dB at $St=0.3$. Thanks to the dipolar nature of the scattered noise, controlling the noise at $-\pi/2$ causes a global reduction at all angles except a small range of $-15^\circ<\theta<5^\circ$. 

\begin{figure}
\centering
\includegraphics[scale=0.80]{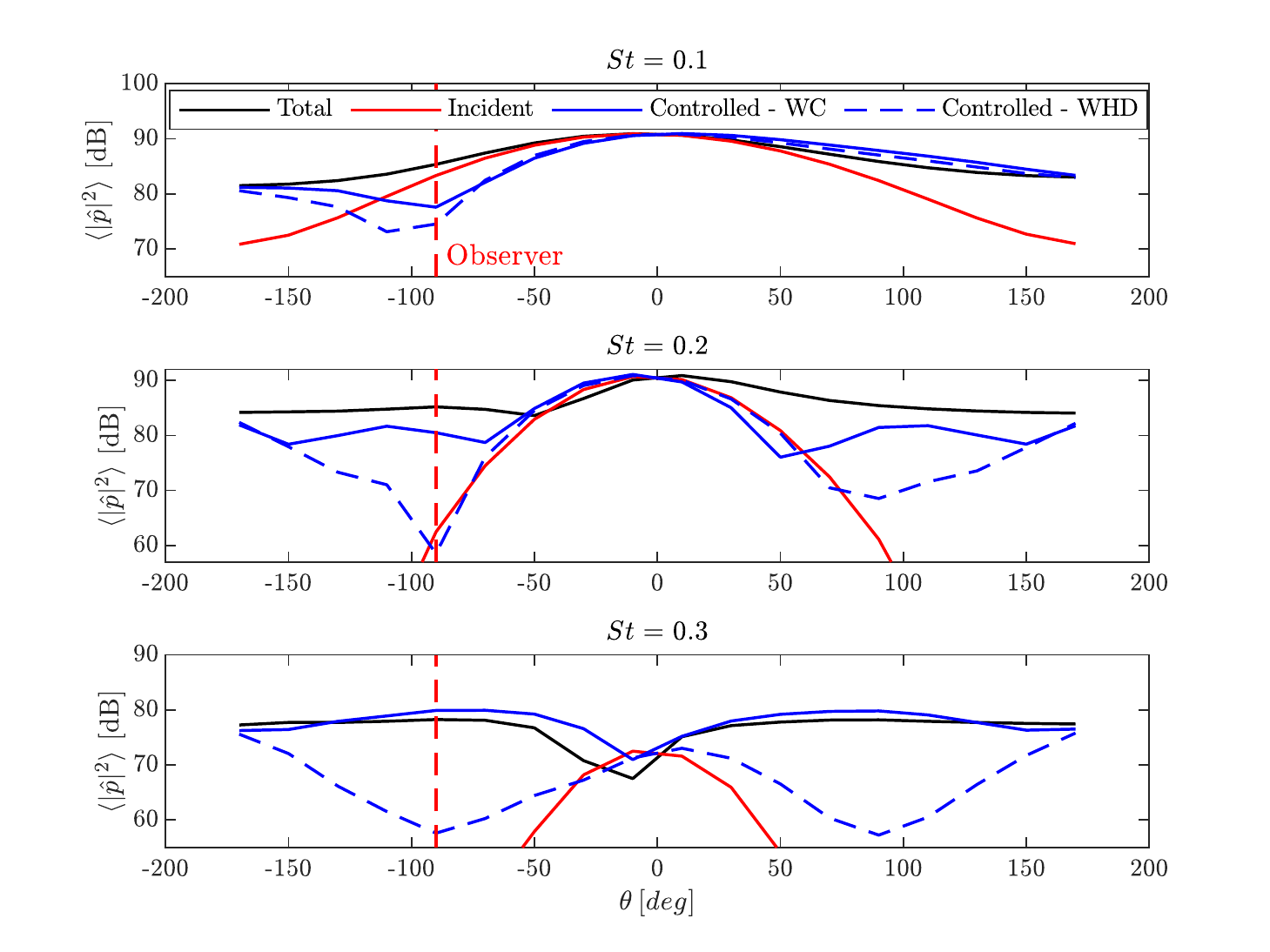}
\caption{The same plot as figure \ref{fig:result2} with sensor position at $x=-5$.}
\label{fig:result5}
\end{figure}

In figure \ref{fig:result5}, we present the same results as in figure \ref{fig:result2}, but with the sensor positioned at $x=-5$. The wave-cancellation approach yields a noise reduction at the control point of around 5 dB at the peak frequencies $St=0.1$ and 0.2 while it causes a slight increase in noise generation at $St=0.3$. The optimal causal control on the other hand provides a 10 dB reduction at $St=0.1$ and an over 20 dB reduction at $St=0.2$ and 0.3. At $St=0.2$, the directivity matches that of the incident source within $-90^\circ<\theta<70^\circ$, indicating that the scattered sound, i.e., the effect of the infinite plate, is entirely cancelled in this directivity range. The increase in the control performance compared to the previous source position case can be explained by increased signal quality at $x=-5$. It was shown in figure \ref{fig:tgfpfield} that the wavepacket envelope peaks at $x=0$ and extends up to $x=\pm6$. The exponential growth mechanism that exists in this region suggests that the disturbances at $x=-5$ have significantly larger amplitude compared to the disturbances at $x=-8$, leading to an improved signal-to-noise ratio given that we assume a constant-level measurement noise in the entire domain.

\begin{figure}
\centering
\includegraphics[scale=0.80]{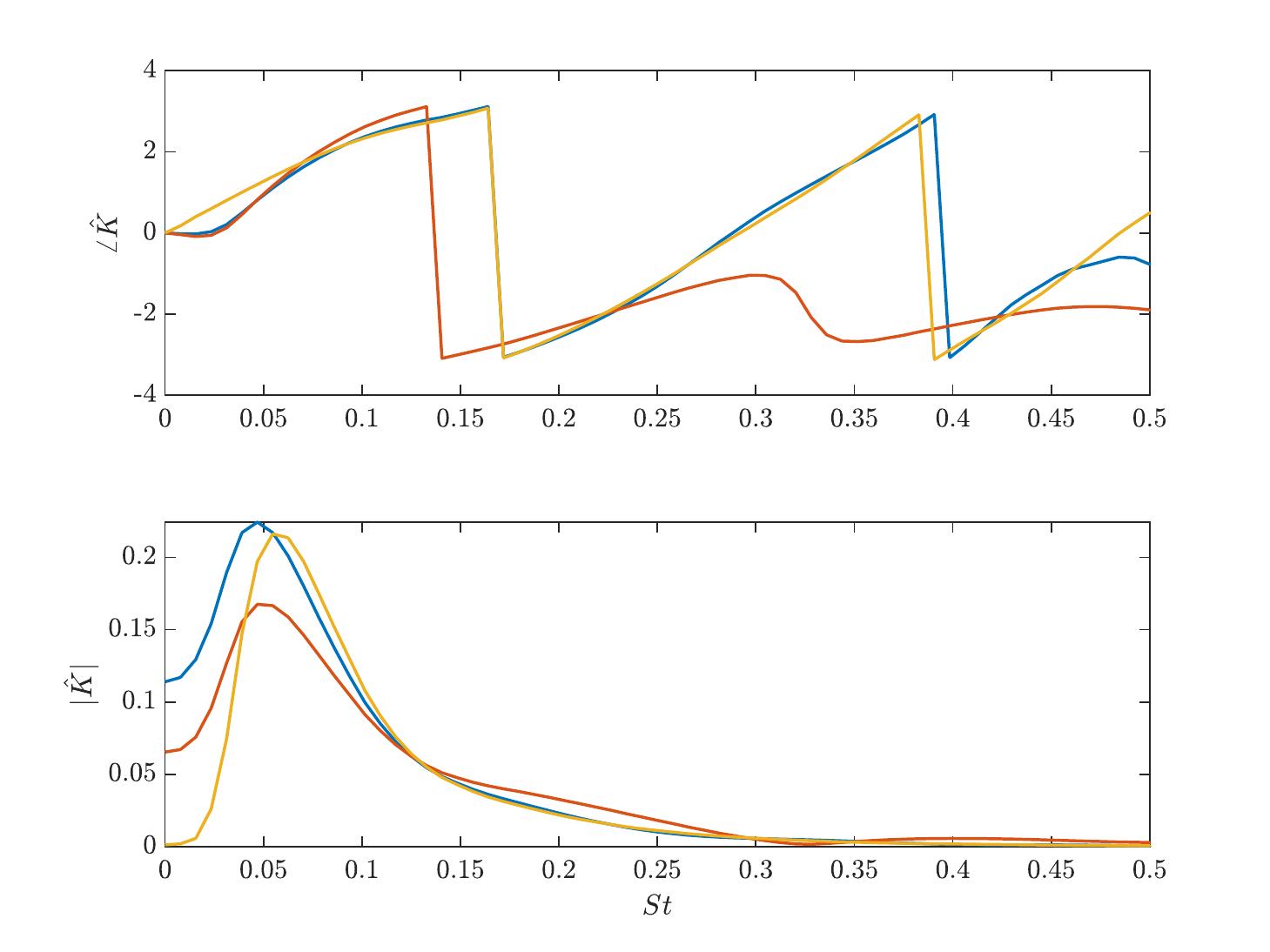}
\caption{Phase (top) and amplitude (bottom) of the control kernels obtained using wave-cancellation (yellow), non-causal (orange) and optimal causal (blue) approaches with sensor position at $x=-5$. Top: phase, bottom: amplitude}
\label{fig:kernelf5}
\end{figure}

The difference in the performance of the control methods with sensor position at $x=-5$ can be investigated by comparing the FTs of the two corresponding kernels against that of the non-causal kernel. The amplitude and phase comparisons are given in figure \ref{fig:kernelf5}. The amplitude spectrum of the kernel obtained from the WC method is similar to the non-causal kernel, where the difference is within $\sim30\%$ for the frequency range $0.05<St<0.4$. Its phase on the other hand starts significantly deviating from that of the non-causal kernel beyond $St=0.3$, which causes the control to be suboptimal at high frequencies leading to the increase in noise as seen in figure \ref{fig:result5}. Contrarily, the amplitude and phase spectra of the kernel obtained from optimal causal control match with good accuracy the non-causal kernel, yielding optimal noise reduction.

\begin{figure}
\centering
\includegraphics[scale=0.80]{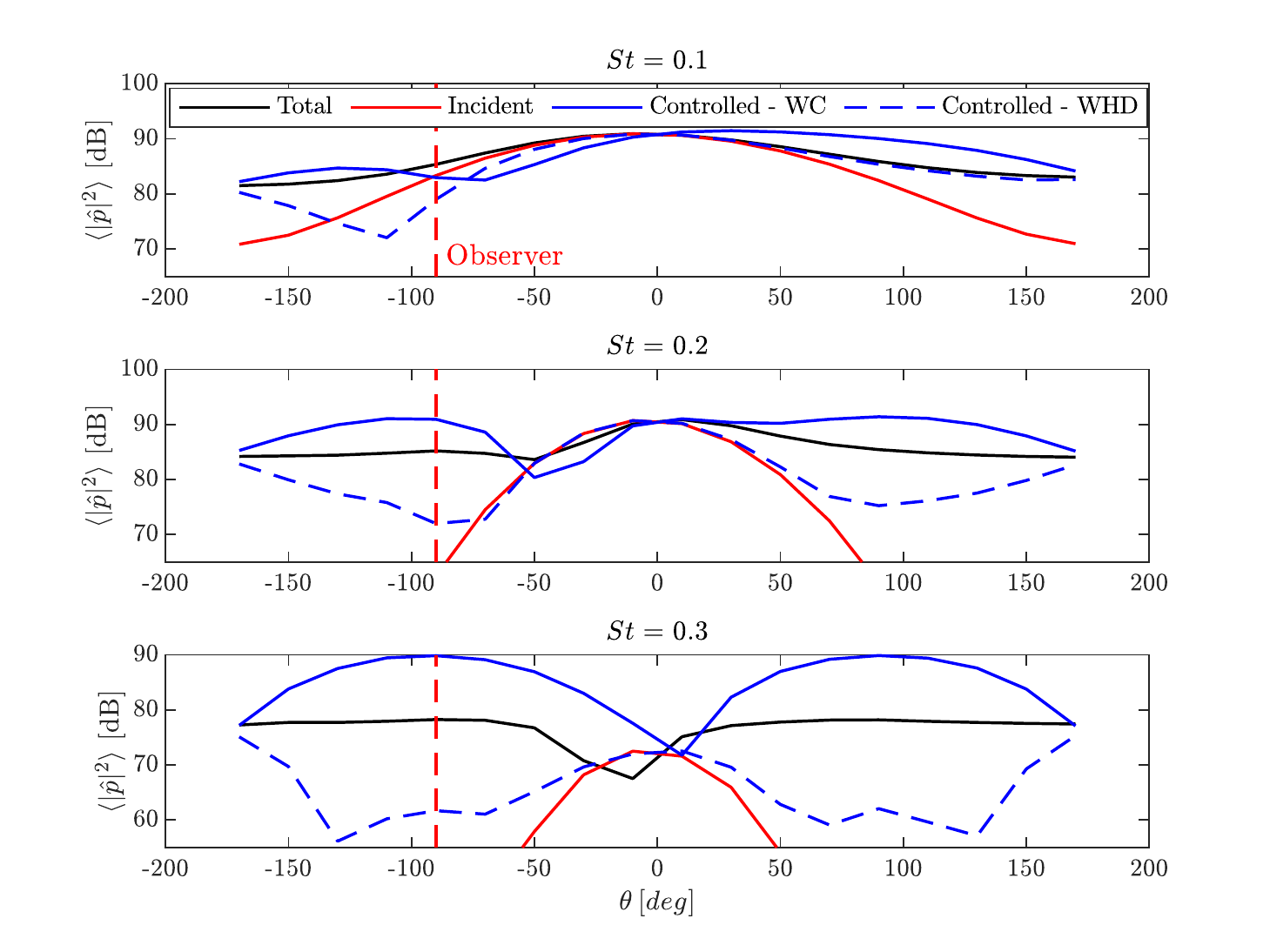}
\caption{The same plot as figure \ref{fig:result2} with sensor position at $x=-2$.}
\label{fig:result8}
\end{figure}

\begin{figure}
\centering
\includegraphics[scale=0.80]{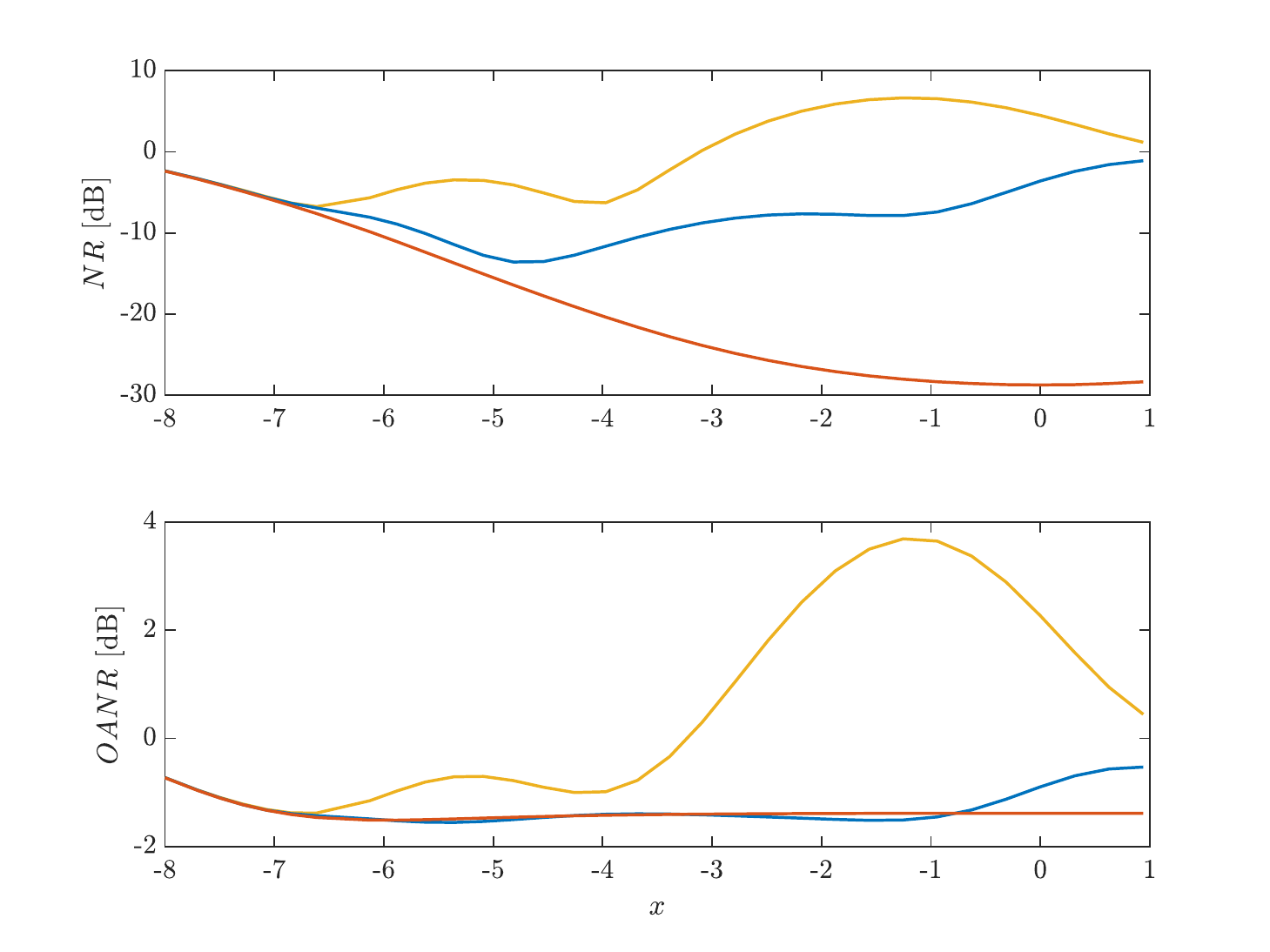}
\caption{The overall noise reduction obtained using \eqref{eq:wckernel} (yellow), \eqref{eq:knc} (orange) and \eqref{eq:ockernel} (blue) at the observer (top) and globally (bottom).}
\label{fig:oanr}
\end{figure}

We now consider the more realistic sensor position at $x=-2$. The directivity plots for this case are shown in figure \ref{fig:result8}. We see that the WC method yields a 3 dB noise reduction at $St=0.1$ while causing more increase on the shielded side. Beyond this frequency, the method  performs significantly worse than the uncontrolled case, causing an increase in noise up to 12 dB. The optimal causal control, once again, provides considerable noise reduction at the observer and also at the rest of the domain. However, the noise reduction achieved is about 5 dB less compared to the previous sensor position, despite the improved signal-to-noise ratio thanks to further amplified perturbations at this position. These results suggest that there exists an optimal sensor position leading to maximum noise reduction, and beyond this point, we expect a worsening of the control performance. This can be investigated by calculating the noise reduction given by the energy ratio at the observer point as
\begin{align}
NR =\frac{\int_0^\infty \langle \lvert \hat{\mathbf{z}}_c\rvert^2\rangle d\omega}{\int_0^\infty \langle \lvert \hat{\mathbf{z}} \rvert^2\rangle d\omega},
\end{align}
where $\hat{\mathbf{z}}_c$ denotes the acoustic pressure at the observer in the controlled case. We calculate the noise reduction using both approaches for different sensor positions in the range $x\in[-8,1]$ and compare the results against that of the non-causal control in the top plot of figure \ref{fig:oanr}. We see that causality starts affecting the performance of the wave-cancellation method for sensor positions downstream of $x=-6.5$, while the decay in the performance starts at sensor positions downstream of $x=-4.5$ for the optimal causal control. In the more realistic case of sensor position around $x=-2$, noise reduction is $8$ dB. The non-causal control, which uses the kernel given by \eqref{eq:knc} without truncating the non-causal part, performs the best when the sensor is positioned at $x=0$, where the wavepacket reaches its maximum amplitude yielding the highest signal-to-noise ratio. 

We evaluate the performance of the control strategy we adopted in this study, which consists of using a dipolar actuator at the TE of the semi-infinite plate, by calculating the overall noise reduction ratio in the entire domain given by 
\begin{align}
OANR =\frac{\int_{-\pi}^{\pi}\int_0^\infty \langle\lvert\hat{\mathbf{z}}_c\rvert^2\rangle d\omega d\theta}{\int_{-\pi}^{\pi}\int_0^\infty \langle\lvert\hat{\mathbf{z}}\rvert^2\rangle d\omega d\theta}.
\end{align}
The results are illustrated in the bottom plot of figure \ref{fig:oanr}. Once again, $OANR$ reaches its minimum for the WC method when the sensor is positioned at $x=-6.5$, leading to a $1.4$ dB reduction in the overall noise. The minimum $OANR$ is achieved with sensor position at $x=-1.5$ for the optimal causal control, leading to an overall noise reduction of $1.5$ dB. Given these results, we can conclude that the optimal causal control can provide substantial noise reduction in a realistic sensor configuration, assuming that an ideal dipolar actuator is available.

\section{Conclusions} \label{sec:conc}
We presented a proof of concept for controlling installation noise in aircraft using a model problem. The model is based on the idea that the installation noise, which has a dipolar directivity \citep{piantanida2016scattering} can be controlled using an actuator at the trailing edge (TE) that can vibrate in the vertical direction, and thus, act as a dipolar source to cancel the installation noise. Assuming that such a control is possible, the problem of feeding the controller with real-time data that would cancel the stochastic noise generated by jet-flap interaction is based on the following hypothesis: the dominant cause of installation noise is the interaction between the wavepackets that appear in the shear layer due to the convective instability of jets, and the trailing edge of the flaps. It was shown by \citet{cavalieri_jfm_2013} that these wavepackets can be modelled using linear solutions of parabolised stability equations (PSE), which yields a rank-1 model, and an accurate prediction of installation noise is then possible using the PSE model \citep{faranosov2019modeling}. As the model is rank-1, the wavepacket can be predicted in real-time via a single-point measurement of the fluctuations in the shear layer, which then opens the possibility of providing input to the controller in real-time. We tested in this study the limits of such a control strategy in a realistic configuration in terms of causality constraints. 

We designed a model that involved a stochastic wavepacket model that is based on the Ginzburg-Landau (G-L) equation and to represent the jet dynamics, a semi-infinite plate to replace the wing flap, and a point dipole placed at the TE of the semi-infinite plate to mimic the actuator. The acoustic propagation is calculated using a tailored Green's function for a semi-infinite plate \citep{fwilliams_jfm_1970}. The observer is located beneath the plate at $90^\circ$ with respect to the TE of the plate, which corresponds to an observer in the ground direction in the actual installation noise problem. The parameters of the G-L problem are set such that the noise generated by the wavepacket has similar characteristics to that of an actual jet at low Mach number, having a dipolar directivity and peaking at Strouhal number $St\approx0.2$ assuming a characteristic length, which is the nozzle diameter in the real problem, of 1. The effect of the sensor position on the causality of the problem is demonstrated. A realistic sensor position that is $\sim2D$ upstream of the flap TE is shown to yield a non-causal control problem. We adopted two control methods: the wave-cancellation (WC) method and optimal causal (OC) control. In the WC method, causal control is achieved by truncating the non-causal part of the kernel. This truncation, however, causes the resulting control kernel to be sub-optimal. In optimal causal control, causality is enforced as a constraint. The two methods yielded similar performance when the sensor is positioned sufficiently upstream of the TE, yielding a causal control problem, albeit not practical considering an actual jet-flap configuration. At the more realistic sensor position of $x=-2$ with the TE at $x=0$, the WC method caused the noise level at the observer to increase by 6 dB, and hence, has proven useless. The OC control on the other hand yielded nearly 8 dB noise reduction at the observer for the same configuration. The global noise reduction was 1.5 dB, which demonstrated that with the control concept adopted, it is possible to achieve significant noise control at a target point without globally generating extra noise. 

The control performance achieved in the model problem is dependent on several factors. The net amount of noise reduction is inversely proportional to the level of measurement noise, which is expected to be different in the actual implementation than the value we assigned in this study. Nonlinearities in an actual jet can also have a negative impact on the noise control approach we adopt, which is linear. Another important factor can be the performance of a real actuator, which might have a limited frequency response or a directivity differing from that of a dipole. We consider here an ideal control problem, and therefore, the control performance achieved in this study demonstrates the upper bound of what one can achieve in a real implementation.  

\backmatter

%
%

\begin{appendices}

\section{Solution of the Wiener-Hopf problem} \label{secA1}
To solve the Wiener-Hopf problem defined in \eqref{eq:wh}, we define two Wiener-Hopf factorisations
\begin{align} \label{eq:whfac1}
\hat{\mathbf{D}}&=\hat{\mathbf{D}}_-\hat{\mathbf{D}}_+,\\
\hat{\mathbf{D}}&=(\hat{\mathbf{D}})_-+(\hat{\mathbf{D}})_+, \label{eq:whfac2}
\end{align}
that are multiplicative and additive, respectively, for a given matrix $\hat{\mathbf{D}}$. Here, the subscripts $-$ and $+$ denote being regular in the lower and upper complex plane, i.e., being entirely causal and non-causal, respectively. For a Wiener-Hopf problem given in the form
\begin{align} \label{eq:whgen}
\hat{\mathbf{D}}\hat{\mathbf{K}}_+\hat{\mathbf{E}} = \hat{\mathbf{\Lambda}}_- + \hat{\mathbf{F}},
\end{align}
the solution for the causal part is given by
\begin{align} \label{eq:ockernelap}
\hat{\mathbf{\Gamma}}_+=\hat{\mathbf{D}}_+^{-1}\big(\hat{\mathbf{D}}_-^{-1}\hat{\mathbf{F}}\hat{\mathbf{E}}_-^{-1}\big)_+\hat{\mathbf{E}}_+^{-1}.
\end{align}
Then, the causal gain matrix in \eqref{eq:wh} can be obtained by setting $\hat{\mathbf{D}}\triangleq\hat{\mathbf{G}}_a^H\hat{\mathbf{G}}_a^{ }+\mathbf{Q}$, $\hat{\mathbf{E}}\triangleq\hat{\mathbf{P}}_{yy}$ and $\hat{\mathbf{F}}\triangleq-\hat{\mathbf{G}}_a^H\hat{\mathbf{P}}_{zy}^{ }$ and substituting these into \eqref{eq:whgen}. We refer the reader to \citet{martini_jfm_2022} for details about how to achieve the Wiener-Hopf factorisations given in \eqref{eq:whfac1} and \eqref{eq:whfac2}, where an efficient matrix-free method based on Hilbert transform is described to perform the multiplicative factorisation. The additive factorisation can be achieved by taking the inverse Fourier transform of $\hat{\mathbf{D}}$ in \eqref{eq:whfac2}, splitting the resulting time-domain kernel $\mathbf{D}$ into two parts such that $\mathbf{D}_-(t<0)=0$ and $\mathbf{D}_+(t>0)=0$, and finally taking the Fourier transforms of $\mathbf{D}_-$ and $\mathbf{D}_+$ to obtain $\hat{\mathbf{D}}_-$ and $\hat{\mathbf{D}}_+$, respectively. 

\section{Spectral proper orthogonal decomposition}\label{secA2}

SPOD \citep{towne_jfm_2018} of a discrete-in-space stochastic variable $\hat{\mathbf{q}}$ can be achieved by calculating the CSD matrix $\hat{\mathbf{P}}_{qq}=\langle\hat{\mathbf{q}}\hat{\mathbf{q}}^H\rangle $ as in \eqref{eq:welch} and then solving the eigenvalue problem
\begin{align} \label{eq:eigint}
\hat{\mathbf{P}}_{qq}\mathbf{W}{\bm{\psi}}={\lambda}{\bm{\psi}},
\end{align}
where ${\bm{\psi}}$ and ${\lambda}$ denote the eigenvector and the eigenvalue, respectively, and $\mathbf{W}$ is a matrix to compute the energy norm, such that the energy of a given stochastic variable $\hat{{\xi}}(x)$ is calculated using the discretised vector $\hat{\bm{\xi}}$ as
\begin{align} \label{eq:enorm}
\langle\hat{\bm{\xi}}^H\mathbf{W}\hat{\bm{\xi}}^H\rangle =\langle\int_S\hat{{\xi}}^*(x)W\hat{{\xi}}(x)dS\rangle ,
\end{align}
where $S$ denotes the domain $\hat{{\xi}}$ is defined, and $W$ denotes the energy norm, which is chosen as 1 for this study. The SPOD modes are then obtained via the eigendecomposition
\begin{align}
\mathbf{W}^{1/2}\hat{\mathbf{P}}_{qq}{{}\mathbf{W}^{1/2}}^H=\tilde{\mathbf{\Psi}}\mathbf{\Lambda}\tilde{\mathbf{\Psi}}^H,
\end{align}
where $\mathbf{\Lambda}$ is a diagonal matrix containing the eigenvalues and $\tilde{\mathbf{\Psi}}\triangleq[\tilde{\bm{\psi}}^{(1)}\,\tilde{\bm{\psi}}^{(2)}\,\cdots]$ denotes the matrix containing the eigenvectors. The eigenvectors given in \eqref{eq:eigint} that are orthogonal with respect to the norm defined in \eqref{eq:enorm} are then calculated as
\begin{align}
{\bm{\psi}}^{(i)}=\mathbf{W}^{-1/2}\tilde{\bm{\psi}}^{(i)},
\end{align}
where the superscript $(i)$ denotes the $i$th eigenvector. The optimal SPOD mode is defined as the eigenvector corresponding to the largest eigenvalue.




\end{appendices}


\section*{Declarations}

\bmhead{Funding}

This work has received funding from the European Union's Horizon 2020 research and innovation programme under grant agreement No 861438. U.K. has received funding from TUBITAK 2236 Co-funded Brain Circulation Scheme 2 (Project No: 121C061).

\bmhead{Author contributions}

All authors contributed to the study conception and design. Ugur Karban and Eduardo Martini developed the codes for the analysis. Ugur Karban drafted the initial version which was edited by Eduardo Martini and Peter Jordan. All authors read and approved the final manuscript.

\bmhead{Competing interests}

The authors have no competing interests to declare.

%
%
%

\bibliography{biblio}


\end{document}